\documentclass[preprint]{elsarticle}
\usepackage{lineno}

\usepackage{amsmath}
\usepackage{subfigure}
\usepackage{float}
\usepackage{graphics}
\usepackage{setspace}
\usepackage{algpseudocode,algorithm}
\usepackage{algorithm}
\usepackage{tikz}
\usepackage{pgf,tikz}
\usepackage{mathrsfs}
\usetikzlibrary{shapes,arrows}
\usepackage{subfigure}
\usepackage{pgfplots}
\usepackage{booktabs}
\usepackage{url}
\usepackage{listings}
\usepackage{color}
\usepackage{courier}
\usepackage[autostyle]{csquotes}
\newtheorem{theorem}{Theorem}[section]

\usepackage{amssymb}

\journal{J. Comput. Phys. }

\begin{document} 

\begin{frontmatter}
	
\title{Binarized octree generation for Cartesian adaptive mesh refinement around immersed geometries \tnoteref{}}
	
	\author{Jaber J. Hasbestan\fnref{myfootnote}}
	\fntext[myfootnote]{Postdoctoral Research Associate, jaber@pitt.edu}
	
	\author{Inanc Senocak\fnref{myfootnote1}}
	\fntext[myfootnote1]{Associate Professor, senocak@pitt.edu}

	\address{Department of Mechanical Engineering and Materials Science,\\ University of Pittsburgh, Pittsburgh, PA 15261, USA}
	
\begin{abstract}
We revisit the generation of balanced octrees for adaptive mesh refinement (AMR) of Cartesian domains with immersed complex geometries. In a recent short note [Hasbestan and Senocak, J. Comput. Phys. vol. 351:473-477 (2017)], we showed that the data-locality of the Z-order curve in hashed linear octree generation methods may not be perfect because of potential collisions in the hash table. Building on that observation, we propose a binarized octree generation method that complies with the Z-order curve exactly. Similar to a hashed linear octree generation method, we use Morton encoding to index the nodes of an octree, but use a red-black tree in place of the hash table. Red-black tree is a special kind of a binary tree, which we use for insertion and deletion of elements during mesh adaptation. By strictly working with the bitwise representation of the octree, we remove computer hardware limitations on the depth of adaptation on a single processor. Additionally, we introduce a geometry encoding technique for rapidly tagging the solid geometry for refinement. Our results for several geometries with different levels of adaptations show that the binarized octree generation outperforms the linear octree generation in terms of runtime performance at the expense of only a slight increase in memory usage. We provide the current AMR capability as open-source software.
\end{abstract}


\end{frontmatter}
\section{Introduction}
Mesh adaptation is a powerful technique in the solution of multiphysics problems governed by coupled partial differential equations (PDEs). By refining the spatial resolution where needed, researchers can realize substantial savings in computer memory and solution turnaround time. Unlike the solution algorithm, mesh adaptation is not governed by a set of governing equations. Therefore, different approaches and strategies have been proposed for mesh adaptation \cite{babushka1983adaptive}. In this study our focus is on the so-called structured adaptive mesh refinement (AMR) method. \citet{berger1984adaptive} pioneered the use of patch-based adaptation for the solution of hyperbolic PDEs with refinement in both space and time. \citet{berger_colella_1989} further improved the AMR algorithm and used an estimate of the local truncation error to guide the refinement decisions. Although these pioneering works were restricted to an explicit finite-difference method, they are one of the early works that introduced efficient use of data-structures in scientific computing and the strategy developed in these works is often referred to as \textit{Berger-Oliger-Colella adaptive refinement}. This unique patch-based strategy has been adopted in Boxlib \cite{boxlib}, PARAMESH \cite{macneice2000paramesh}, AMREx \cite{AMREx}, Chombo \cite{adams2014chombo}, ENZO \cite{bryan2014enzo}, SAMRAI \cite{samrai, samrai_article} and AMRClaw \cite{Berger_LeVeque_1998} packages. In a patch-based AMR, a hierarchy of patches are nested within a coarse level mesh that occupies the entire domain. A refinement ratio is assigned between successive patches. During refinement, patches are tagged based on a specified criterion for refinement. Ghost-cells are used at patch boundaries to compute the solution across the patches. 
  
Implementation of an AMR algorithm is major software development effort. It is common to refer to the end product as an AMR infrastructure. In building this software infrastructure, the design goal is to introduce an abstraction for the meshing part such that developers of the solution algorithm need not know the details of the computer implementation. For static-mesh problems, the efficiency of the AMR infrastructure may not be critical. But for large-scale problems in which the refinement criteria is dynamic, time spent on meshing, while the simulation is evolving, can be prohibitively expensive. Additionally, the memory hierarchy on modern computing hardware favors computer implementations that exploit data-locality and use less memory. It is worth mentioning that these performance issues related to adaptive meshing have been the focus of several studies in the computer graphics field, driven by real-time applications \cite{lewiner2010fast}. There are a number of open-source packages that provide AMR capabilities for meshing and solving PDEs. \citet{dubey_etal_2014} survey a representative set of block-structured AMR infrastructures.

An alternative strategy to the patch-based AMR approach is a tree-based strategy. Octree is a hierarchical data structure that has been the subject of many studies \cite{meagher1982geometric, samet1984quadtree}. The tree structure has also been adopted in the solution of partial differential equations (PDEs) with adaptive resolutions. In its classical representation, an octree keeps track of parent-child relations with pointers. In the three-dimensional case, eight pointers are needed to store the neighbors.

\citet{sundar2008bottom} developed parallel algorithms to generate balanced linear octrees. A linear octree is a representation of an octree in which a linear array is used instead of a tree data structure to store the leaf nodes \cite{gargantini1982a,gargantini1982b}. 
It is common to use space-filling curves to represent linear octrees. Sundar et al. used Morton encoding as one of the many ways to generate a space-filling curve because of its ease of computer implementation. Their method addressed the so-called \textit{ripple-effect} associated with balancing the octree.  

\citet{burstedde2011p4est} pursued the linear octree representation approach described in \citet{sundar2008bottom} and developed \texttt{p4est}, which is an AMR package for multiple connected octrees. Authors refer to these connected octrees as \textit{forest-of-octrees}. An encoding scheme was introduced to enable inter-octree connectivity. Octrees and octants were numbered using a space-filling Z-order curve, which was effective in parallel load balancing. \texttt{p4est} has been shown scale across many processors. In one exercise, a mesh with $5.13\times10^{11}$ elements was created using 220,320 processes. \citet{isaac2012low} introduced recursive algorithms for this forest-of-octrees approach to traverse a linear octree and further improve the parallel scalability of \texttt{p4est} aimed at improving the 2:1 balance algorithm. 

Another approach is the building-cube method (BCM) \cite{ishikawa2011large,ishida2008efficient}, which is a combination of explicit tree and unstructured grid methodology. In BCM, each cube has its own equally-spaced Cartesian grid. All cubes have the same number of Cartesian mesh. Therefore, the resolution is determined by the level of adaptation of the cube.  

Having a balanced tree is important in numerical solution of PDEs as it enables one to gradually traverse from low-resolution to high-resolution without significantly degrading numerical accuracy, especially for simulation of multi-scale problems such as turbulent flows. A balanced-tree is also needed to enable a generic implementation of numerical schemes for the solution of PDEs in a code because coarse to fine blocks obeys the same rule. The common approach to enforce a balanced tree is to confine the maximum level of the element at each face to $2^{d-1}$ where $d$ is the dimension of the computational space (i.e. 2 or 3). The difficulty of the 2:1 balancing algorithm emanates from the fact that the domain of influence of any element under consideration for refinement might extend beyond the immediate neighboring elements. This issue has been coined as the \textit{ripple effect} in the literature \cite{sundar2008bottom}. Different methods have been proposed to improve the efficiency of enforcing the 2:1 balance condition.

\citet{tu2004balance} demonstrate 2:1 balanced refinement of octrees. Their solution consists of two steps; balance by parts and prioritize the\textit{ ripple effect}. The key idea in their approach is to bulk load most of the data on memory and enforce the balance locally. They load a small region to a pointer-based  (sub)octree  in memory and resolve the 2:1 constraint and the ripple effect  without worrying about octants outside of the region. This is due to the fact that most impact  caused  by  a  tiny  octant tagged to be refined is localized  in  a  small region. Their balance-by-parts algorithm  is described as follows.  Imagine  a moving  window  inside  the 3D domain,  at  any given  moment,  the octants that reside in this window are retrieved  from  disk  and cached  in a (sub)octree. Then,  they  adjust  the  data  structure  to  enforce  2-to-1  constraint in  memory,  the  content  on  disk  is updated  accordingly.  

\textit{Ripple effect} is even more challenging in distributed systems due to the complicated communication pattern. \citet{sundar2008bottom} proposed a bottom-up approach to enforce the balance condition. This approach eliminates the need for synchronization in parallel implementation. \citet{isaac2012low} improved the balance algorithm in the \texttt{p4est} package. In their approach, they first reduce the input set to a compressed equivalent and iteratively add a sparse set of octants to that list to enforce the balance. 

Aside from solving the PDEs, octrees are extensively used in computer graphics. Unlike in the solution of PDEs, balanced trees are not required in computer graphics applications, and the desired resolution can be much lower than resolutions needed for turbulent flow simulations. Driven by interactivity requirements of computer games and ray tracing applications, the implementation of octree on central processing units (CPUs) or graphics processing units (GPUs) can be highly optimized. Morton codes are used in building octrees, mapping multi-dimensional data to one dimension and thereby enhancing data-locality. Use of a hash table with Morton encoding is common in generating the octrees. \citet{meagher1982geometric} utilized octree data structure for modeling the three-dimensional complex geometries including geometries with holes and disjointed parts and sculptured surfaces. \citet{frisken2002simple} demonstrate efficient methods for tree structure traversals. 

\citet{lewiner2010fast} developed a fast pointerless method to generate dual octrees. They proposed strategies for dual generation of static or dynamic pointerless octrees, where memory usage is lowered by several factors compared to the usual recursive generation methodologies. Note that pointerless representation is another terminology for hashed linear octrees. However, their study is not related to solving PDEs and hence, the generated mesh is not necessarily balanced. Lewiner et al. provide a good review on different approaches to octree generation in the same work.

\citet{schwarz2010fast} describe octree implementation on GPUs.
\citet{baert2014out} uses bit-interleaving in constructing the Morton code to map every voxel center to a one-dimensional space to generate out-of-core octrees for a Cartesian mesh, the method presented there is not directly
applicable to AMR, as they use integer indices in the mapping. The bit-interleaving introduced in that study uses fixed point arithmetic to store the center of the cube. Bit-interleaving is also introduced by \citet{NVIDIA}. The problem with these approaches for solving PDEs is that one needs to always scale the geometry to unit length and extracting the connectivity is not trivial. 

Different versions of space-filling curves have been used for partitioning the octrees \cite{campbell2003dynamic,flaherty1997adaptive,brunet1990solid,vo2012simple}.
Z-order curve, also referred to as Morton code \cite{morton1966computer}, is often preferred because of its ease of computer implementation \cite{sundar2008bottom}. An important feature of using Morton code is that it ensures that the elements with closer Morton  codes are stored close to each other in the computer memory. This feature makes the search for a given key efficient, since one only needs to look for the elements close to the current key. 

The method used to search for a key is also critical for performance. In a recent short note \cite{hasbestan2017shortnote}, we have shown that a red-black tree helps preserve the Z-order curve exactly, whereas the Z-order can be broken because of collisions in a hashed linear octree. A red-black tree also supports deep-level mesh adaptation for extreme-scale problems because the method is based on bitwise comparisons only. 

Building on those useful features of the red-black tree data structure, we present a fast, low-storage AMR algorithm that complies with the Z-order curve strictly. Our presentation includes the algorithms for 2:1 balance, geometry encoding for immersion of complex geometries, and the tricks and trades of computer implementation that exploits bitwise representation heavily. In our approach, we do not store the neighborhood connectivity explicitly. To the best of our knowledge, our open-source software, which we call \texttt{rebl-AMR}, is the first Cartesian adaptive mesh generation package that make use of the fundamental red-black tree data structure at its core. We present numerous examples to demonstrate the performance of red-black tree version and compare its computational performance to a hashed linear octree. For completeness, we compare the performance results from both methods against a graph-based unstructured mesh approach as well. The present rebl-AMR  \cite{rbAMR} package is available as open-source software.



\section{Octree representations} 
An octree is constructed by isotropic refinement where a rectangular volume is subdivided in half in $x$, $y$ and $z$ directions.
Refinement of each element generates eight more elements in three dimensions. We refer to these elements as \textit{siblings}. 
Complicated geometries can be represented using octrees by recursively subdividing the elements.
Different types of data structures have been used to store the tree data. 
The selection of the data structures is a crucial step in octree generation algorithms because it will impact the execution time as well as memory usage. For dynamic adaptive mesh generation problems where the geometry deforms or moves in time, it is desirable to have constant-in-time insertion and removal of mesh elements.

One of the early approaches in octree generation is the use of parent-child relation in creating a so-called explicit octree, where use of pointers
to children is standard. This approach is the classical way of storing the tree and uses eight pointers for octree representation  \cite{drozdek2012data}. A sample four-level tree is presented in Fig. \ref{fig:sample}. The number of pointers needed to represent a large adaptively refined mesh would be unnecessarily large as well and would also complicate the computer implementation. Pointerless implicit trees have been proposed as an alternative to the explicit tree representation \cite{gargantini1982a, gargantini1982b}.

We present three methods to generate octrees. The first method is the hashed linear octree generation. Although this method is very efficient, there is no guarantee to preserve the Z-order curve once it is used in conjunction with a hash table because of collision, which we have demonstrated recently \cite{hasbestan2017shortnote}. The second method is unique to the present work, which we call it the \textit{binarized} octree generation. Similar to a hashed linear octree generation method, we represent octree nodes on a Z-order curve and use a red-black tree for element search, insertion and deletion operations. A red-black is a special kind of binary-tree and bitwise representation is used to implement the Z-order curve. Because we use a binary tree data structure to store the leaves of the octree instead of an array, we refer to this approach as \textit{binarized octree generation}. Compared to the classical approach to generate an octree, both the hashed linear octree and the binarized octree use much less memory.

In the third approach, we generate an octree as an unstructured mesh.
The connectivity is generated for every element using a graph-based approach. This method updates the connectivity directly without the need to actually store the tree information explicitly. The unstructured mesh approach is provided for completeness. We describe that approach in \citet{hasbestan2017parallel} and it is not the main focus of the current work.

Z-order curve, also referred to as the Morton code, is common to both the linear and binarized octree generation methods. Therefore, we first explain the construction of Morton codes and how we manipulate them for neighborhood connectivity information.

\subsection{Morton code generation}
\label{subsection:morton}
Morton encoding is a mapping from a multi-dimensional space to one dimension \cite{morton1966computer}. When generating a Morton code, first, a bit code is constructed for every node. This node is then converted to an integer, if needed. The nodes, once laid out, follow a Z-order curve, which enhances data-locality.

\begin{figure}[h]
	\pagestyle{empty}
	\definecolor{zzccff}{rgb}{0.6,0.8,1.}
	\definecolor{ccccff}{rgb}{0.8,0.8,1.}
	\definecolor{qqzzff}{rgb}{0.,0.6,1.}
	\definecolor{zzffff}{rgb}{0.6,1.,1.}
	\definecolor{ffcctt}{rgb}{1.,0.8,0.2}
	\definecolor{ffffww}{rgb}{1.,1.,0.4}
	\definecolor{ccwwqq}{rgb}{0.8,0.4,0.}
	\definecolor{ffzztt}{rgb}{1.,0.6,0.2}
	\definecolor{qqqqff}{rgb}{0.,0.,1.}
	\definecolor{xdxdff}{rgb}{0.49019607843137253,0.49019607843137253,1.}
	\definecolor{uuuuuu}{rgb}{0.26666666666666666,0.26666666666666666,0.26666666666666666}
	\begin{tikzpicture}[line cap=round,line join=round,>=triangle 45,x=1.0cm,y=1.0cm,scale=0.9]
	\fill[color=ffzztt,fill=ffzztt,fill opacity=1.0] (0.0014221391516796337,0.006083629927127855) -- (2.0009819567282374,0) -- (3.0001575054017837,0.9999453387351146) -- (0.9978058988834555,0.997663626037123) -- cycle;
	\fill[color=ccwwqq,fill=ccwwqq,fill opacity=1.0] (2.0009819567282374,0) -- (4.001918943793339,0.) -- (5.002681196999808,0.997663626037123) -- (3.0001575054017837,0.9999453387351146) -- cycle;
	\fill[color=ffffww,fill=ffffww,fill opacity=1.0] (3.0001575054017837,0.9999453387351146) -- (5.002681196999808,0.997663626037123) -- (5.999755763967,2.0010438005400126) -- (4.000426146346458,2.0001685504032736) -- cycle;
	\fill[color=ffcctt,fill=ffcctt,fill opacity=1.0] (1.9990247234125438,1.9988824505662217) -- (0.9978058988834555,0.997663626037123) -- (3.0001575054017837,0.9999453387351146) -- (4.000426146346458,2.0001685504032736) -- cycle;
	\fill[color=zzffff,fill=zzffff,fill opacity=1.0] (0.,4.) -- (1.9981224937881954,3.997494616540729) -- (3.0002435479416323,5.002538924153518) -- (0.9999026374089197,4.999593532684708) -- cycle;
	\fill[color=qqzzff,fill=qqzzff,fill opacity=1.0] (1.9981224937881954,3.997494616540729) -- (4.000297776521328,3.999772082539966) -- (4.999128487705536,4.998917966974395) -- (3.0002435479416323,5.002538924153518) -- cycle;
	\fill[color=ccccff,fill=ccccff,fill opacity=1.0] (3.0002435479416323,5.002538924153518) -- (4.999128487705536,4.998917966974395) -- (5.997563193525548,5.997420920679268) -- (3.999734248851152,6.000071907511541) -- cycle;
	\fill[color=zzccff,fill=zzccff,fill opacity=1.0] (0.9999026374089197,4.999593532684708) -- (3.0002435479416323,5.002538924153518) -- (3.999734248851152,6.000071907511541) -- (1.9990247234125438,6.003757748682617) -- cycle;
	\draw (1.214828264277152,0.661198387660524) node[anchor=north west] {000};
	\draw (3.0911358257933452,0.6439845745341681) node[anchor=north west] {100};
	\draw (1.3095043338949415,4.551520154216943) node[anchor=north west] {001};
	\draw (3.1944188108309337,4.5429132476537655) node[anchor=north west] {101};
	\draw (2.3423341842708276,5.6618111008668945) node[anchor=north west] {011};
	\draw (4.304710899985011,5.653204194303717) node[anchor=north west] {111};
	\draw (4.227248661206819,1.5649235767942054) node[anchor=north west] {110};
	\draw (2.187409706714445,1.6423857358628067) node[anchor=north west] {010};
	\draw [->] (-3.5,0.) -- (-2.,0.);
	\draw [->] (-3.5,0.) -- (-2.5,1.);
	\draw [->] (-3.5,0.) -- (-3.5,1.5);
	\draw (5.997563193525548,5.997420920679268)-- (6.000262378487905,4.000467442211195);
	\draw (6.000262378487905,4.000467442211195)-- (5.999755763967,2.0010438005400126);
	\draw (6.000262378487905,4.000467442211195)-- (5.000759660610562,3.0003353420024457);
	\draw (4.000426146346458,2.0001685504032736)-- (5.000759660610562,3.0003353420024457);
	\draw [dash pattern=on 2pt off 2pt] (4.000297776521328,3.999772082539966)-- (6.000262378487905,4.000467442211195);
	\draw [dash pattern=on 2pt off 2pt] (4.000426146346458,2.0001685504032736)-- (3.0001575054017837,0.9999453387351146);
	\draw [dash pattern=on 2pt off 2pt] (3.0001575054017837,0.9999453387351146)-- (3.0005499818746615,3.0011795879820378);
	\draw [dash pattern=on 2pt off 2pt] (5.000759660610562,3.0003353420024457)-- (3.0005499818746615,3.0011795879820378);
	\draw [dash pattern=on 2pt off 2pt] (3.0005499818746615,3.0011795879820378)-- (1.9990247234125438,1.9988824505662217);
	\draw [dash pattern=on 2pt off 2pt] (1.0003239157081802,3.0001945301357686)-- (3.0005499818746615,3.0011795879820378);
	\draw [dash pattern=on 2pt off 2pt] (3.0002435479416323,5.002538924153518)-- (3.0005499818746615,3.0011795879820378);
	\draw [dash pattern=on 2pt off 2pt] (3.0005499818746615,3.0011795879820378)-- (4.000297776521328,3.999772082539966);
	\draw (4.001918943793339,0.)-- (4.000426146346458,2.0001685504032736);
	\draw (1.9990247234125438,1.9988824505662217)-- (4.000426146346458,2.0001685504032736);
	\draw [dash pattern=on 2pt off 2pt] (4.000426146346458,2.0001685504032736)-- (5.999755763967,2.0010438005400126);
	\draw [dash pattern=on 2pt off 2pt] (0.9978058988834555,0.997663626037123)-- (0.0014221391516796337,0.006083629927127855);
	\draw [dash pattern=on 2pt off 2pt] (0.9978058988834555,0.997663626037123)-- (1.0003239157081802,3.0001945301357686);
	\draw [dash pattern=on 2pt off 2pt] (1.9990247234125438,1.9988824505662217)-- (0.9978058988834555,0.997663626037123);
	\draw [dash pattern=on 2pt off 2pt] (3.0001575054017837,0.9999453387351146)-- (0.9978058988834555,0.997663626037123);
	\draw [dash pattern=on 2pt off 2pt] (1.9981224937881954,3.997494616540729)-- (1.0003239157081802,3.0001945301357686);
	\draw (1.9981224937881954,3.997494616540729)-- (1.9990247234125438,1.9988824505662217);
	\draw [dash pattern=on 2pt off 2pt] (1.9990247234125438,6.003757748682617)-- (1.9981224937881954,3.997494616540729);
	\draw (0.,4.)-- (1.9981224937881954,3.997494616540729);
	\draw (1.9981224937881954,3.997494616540729)-- (4.000297776521328,3.999772082539966);
	\draw (4.000297776521328,3.999772082539966)-- (4.000426146346458,2.0001685504032736);
	\draw (4.999128487705536,4.998917966974395)-- (5.997563193525548,5.997420920679268);
	\draw (4.999128487705536,4.998917966974395)-- (5.000759660610562,3.0003353420024457);
	\draw (4.999128487705536,4.998917966974395)-- (4.000297776521328,3.999772082539966);
	\draw (4.999128487705536,4.998917966974395)-- (3.0002435479416323,5.002538924153518);
	\draw (1.9990247234125438,6.003757748682617)-- (0.9999026374089197,4.999593532684708);
	\draw (0.9999026374089197,4.999593532684708)-- (0.,4.);
	\draw (0.9999026374089197,4.999593532684708)-- (3.0002435479416323,5.002538924153518);
	\draw (0.0014221391516796337,0.006083629927127855)-- (2.0009819567282374,0);
	\draw (2.0009819567282374,0)-- (4.001918943793339,0.);
	\draw [dash pattern=on 2pt off 2pt] (2.0009819567282374,0)-- (3.0001575054017837,0.9999453387351146);
	\draw (5.999755763967,2.0010438005400126)-- (5.002681196999808,0.997663626037123);
	\draw (5.000759660610562,3.0003353420024457)-- (5.002681196999808,0.997663626037123);
	\draw (5.002681196999808,0.997663626037123)-- (4.001918943793339,0.);
	\draw [dash pattern=on 2pt off 2pt] (5.002681196999808,0.997663626037123)-- (3.0001575054017837,0.9999453387351146);
	\draw (1.9990247234125438,1.9988824505662217)-- (2.0009819567282374,0);
	\draw [dash pattern=on 2pt off 2pt] (0.9999026374089197,4.999593532684708)-- (1.0003239157081802,3.0001945301357686);
	\draw (0.,4.)-- (-2.845771829521041E-4,1.9991946233956495);
	\draw (-2.845771829521041E-4,1.9991946233956495)-- (0.0014221391516796337,0.006083629927127855);
	\draw (-2.845771829521041E-4,1.9991946233956495)-- (1.9990247234125438,1.9988824505662217);
	\draw [dash pattern=on 2pt off 2pt] (1.0003239157081802,3.0001945301357686)-- (-2.845771829521041E-4,1.9991946233956495);
	\draw (1.9990247234125438,6.003757748682617)-- (3.999734248851152,6.000071907511541);
	\draw (3.999734248851152,6.000071907511541)-- (5.997563193525548,5.997420920679268);
	\draw (3.999734248851152,6.000071907511541)-- (3.0002435479416323,5.002538924153518);
	\draw [color=ffzztt] (0.0014221391516796337,0.006083629927127855)-- (2.0009819567282374,0);
	\draw [color=ffzztt] (2.0009819567282374,0)-- (3.0001575054017837,0.9999453387351146);
	\draw [color=ffzztt] (3.0001575054017837,0.9999453387351146)-- (0.9978058988834555,0.997663626037123);
	\draw [color=ffzztt] (0.9978058988834555,0.997663626037123)-- (0.0014221391516796337,0.006083629927127855);
	\draw [color=ccwwqq] (2.0009819567282374,0)-- (4.001918943793339,0.);
	\draw [color=ccwwqq] (4.001918943793339,0.)-- (5.002681196999808,0.997663626037123);
	\draw [color=ccwwqq] (5.002681196999808,0.997663626037123)-- (3.0001575054017837,0.9999453387351146);
	\draw [color=ccwwqq] (3.0001575054017837,0.9999453387351146)-- (2.0009819567282374,0);
	\draw [color=ffffww] (3.0001575054017837,0.9999453387351146)-- (5.002681196999808,0.997663626037123);
	\draw [color=ffffww] (5.002681196999808,0.997663626037123)-- (5.999755763967,2.0010438005400126);
	\draw [color=ffffww] (5.999755763967,2.0010438005400126)-- (4.000426146346458,2.0001685504032736);
	\draw [color=ffffww] (4.000426146346458,2.0001685504032736)-- (3.0001575054017837,0.9999453387351146);
	\draw [color=ffcctt] (1.9990247234125438,1.9988824505662217)-- (0.9978058988834555,0.997663626037123);
	\draw [color=ffcctt] (0.9978058988834555,0.997663626037123)-- (3.0001575054017837,0.9999453387351146);
	\draw [color=ffcctt] (3.0001575054017837,0.9999453387351146)-- (4.000426146346458,2.0001685504032736);
	\draw [color=ffcctt] (4.000426146346458,2.0001685504032736)-- (1.9990247234125438,1.9988824505662217);
	\draw [color=zzffff] (0.,4.)-- (1.9981224937881954,3.997494616540729);
	\draw [color=zzffff] (1.9981224937881954,3.997494616540729)-- (3.0002435479416323,5.002538924153518);
	\draw [color=zzffff] (3.0002435479416323,5.002538924153518)-- (0.9999026374089197,4.999593532684708);
	\draw [color=zzffff] (0.9999026374089197,4.999593532684708)-- (0.,4.);
	\draw [color=qqzzff] (1.9981224937881954,3.997494616540729)-- (4.000297776521328,3.999772082539966);
	\draw [color=qqzzff] (4.000297776521328,3.999772082539966)-- (4.999128487705536,4.998917966974395);
	\draw [color=qqzzff] (4.999128487705536,4.998917966974395)-- (3.0002435479416323,5.002538924153518);
	\draw [color=qqzzff] (3.0002435479416323,5.002538924153518)-- (1.9981224937881954,3.997494616540729);
	\draw [color=ccccff] (3.0002435479416323,5.002538924153518)-- (4.999128487705536,4.998917966974395);
	\draw [color=ccccff] (4.999128487705536,4.998917966974395)-- (5.997563193525548,5.997420920679268);
	\draw [color=ccccff] (5.997563193525548,5.997420920679268)-- (3.999734248851152,6.000071907511541);
	\draw [color=ccccff] (3.999734248851152,6.000071907511541)-- (3.0002435479416323,5.002538924153518);
	\draw [color=zzccff] (0.9999026374089197,4.999593532684708)-- (3.0002435479416323,5.002538924153518);
	\draw [color=zzccff] (3.0002435479416323,5.002538924153518)-- (3.999734248851152,6.000071907511541);
	\draw [color=zzccff] (3.999734248851152,6.000071907511541)-- (1.9990247234125438,6.003757748682617);
	\draw [color=zzccff] (1.9990247234125438,6.003757748682617)-- (0.9999026374089197,4.999593532684708);
	\begin{scriptsize}
	\draw [fill=uuuuuu] (0.,0.) circle (1.5pt);
	\draw [fill=xdxdff] (4.001918943793339,0.) circle (2.5pt);
	\draw [fill=qqqqff] (6.,2.) circle (2.5pt);
	\draw [fill=qqqqff] (4.,4.) circle (2.5pt);
	\draw [fill=qqqqff] (6.,6.) circle (2.5pt);
	\draw [fill=qqqqff] (2.,6.) circle (2.5pt);
	\draw [fill=xdxdff] (0.,4.) circle (2.5pt);
	\draw [fill=qqqqff] (2.,2.) circle (2.5pt);
	\draw [fill=qqqqff] (1.,3.) circle (2.5pt);
	\draw [fill=qqqqff] (0.0014221391516796337,0.006083629927127855) circle (2.5pt);
	\draw [fill=qqqqff] (-2.845771829521041E-4,1.9991946233956495) circle (2.5pt);
	\draw [fill=qqqqff] (-5.866387819546103E-4,4.000445861107676) circle (2.5pt);
	\draw [fill=qqqqff] (1.0003239157081802,3.0001945301357686) circle (2.5pt);
	\draw [fill=qqqqff] (0.9999026374089197,4.999593532684708) circle (2.5pt);
	\draw [fill=qqqqff] (1.9981224937881954,3.997494616540729) circle (2.5pt);
	\draw [fill=qqqqff] (2.0009819567282374,0) circle (2.5pt);
	\draw [fill=qqqqff] (4.002575467978908,0.0015116386288441998) circle (2.5pt);
	\draw [fill=qqqqff] (5.002681196999808,0.997663626037123) circle (2.5pt);
	\draw [fill=qqqqff] (5.999755763967,2.0010438005400126) circle (2.5pt);
	\draw [fill=qqqqff] (6.000262378487905,4.000467442211195) circle (2.5pt);
	\draw [fill=qqqqff] (4.999128487705536,4.998917966974395) circle (2.5pt);
	\draw [fill=qqqqff] (4.000297776521328,3.999772082539966) circle (2.5pt);
	\draw [fill=qqqqff] (5.997563193525548,5.997420920679268) circle (2.5pt);
	\draw [fill=qqqqff] (3.999734248851152,6.000071907511541) circle (2.5pt);
	\draw [fill=qqqqff] (1.9990247234125438,6.003757748682617) circle (2.5pt);
	\draw [fill=qqqqff] (3.0002435479416323,5.002538924153518) circle (2.5pt);
	\draw [fill=qqqqff] (3.0005499818746615,3.0011795879820378) circle (2.5pt);
	\draw [fill=qqqqff] (5.000759660610562,3.0003353420024457) circle (2.5pt);
	\draw [fill=qqqqff] (4.000426146346458,2.0001685504032736) circle (2.5pt);
	\draw [fill=qqqqff] (1.9990247234125438,1.9988824505662217) circle (2.5pt);
	\draw [fill=qqqqff] (0.9978058988834555,0.997663626037123) circle (2.5pt);
	\draw [fill=qqqqff] (3.0001575054017837,0.9999453387351146) circle (2.5pt);
	\draw [fill=qqqqff] (-3.5,0.) circle (0.5pt);
	\draw [fill=qqqqff] (-2.,0.) circle (0.5pt);
	\draw[color=qqqqff] (-1.9439096947891,0.08883910120919236) node {$X$};
	\draw [fill=qqqqff] (-2.5,1.) circle (0.5pt);
	\draw[color=qqqqff] (-2.4431107891374446,1.0872402625378308) node {$Y$};
	\draw [fill=qqqqff] (-3.5,1.5) circle (0.5pt);
	\draw[color=qqqqff] (-3.4415129778341345,1.5864408432021502) node {$Z$};
	\end{scriptsize}
	\end{tikzpicture}
	\caption{Morton code construction at a single level based on element location in x, y, z directions}
	\label{fig:3D}
\end{figure}
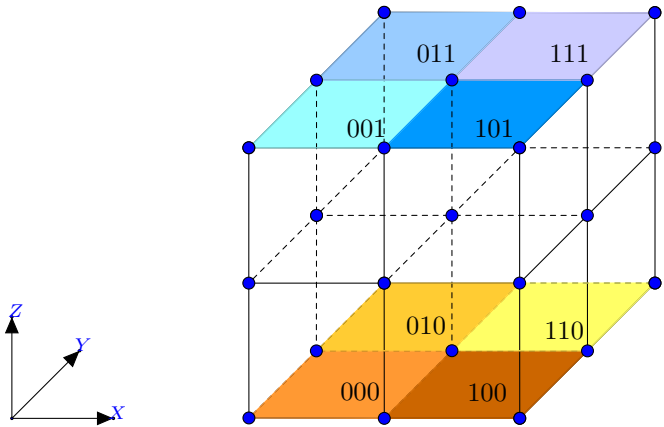

Figure \ref{fig:3D} presents the Morton ordering for a mesh in three dimensions, which is refined one level only. Note that we colored the four lower octants on their bottom surface while coloring the top four octants on their top surfaces to avoid visual clutter. The bit representation on each surface is valid for the individual octant. The Morton code for each octant is constructed by specifying its relative location index in  $x, y, z$ directions in the Cartesian coordinates, respectively.

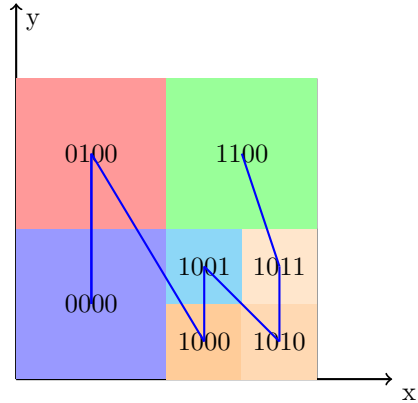
\begin{figure}[h]
	\centering
	\begin{tikzpicture}
	\draw[step=2cm,gray,very thin] (0,0) grid (4,4);
	\draw[step=1cm,gray,very thin] (2,0) grid (4,2);
	\draw[thick,->] (0,0) -- (5.0,0) node[anchor=north west] {x};
	\draw[thick,->] (0,0) -- (0,5.0) node[anchor=north west] {y};;
	\fill[blue!40!white] (0,0) rectangle (2,2);
	\fill[orange!40!white] (2,0) rectangle (3,1);
	\fill[orange!30!white] (3,0) rectangle (4,1);
	\fill[orange!20!white] (3,1) rectangle (4,2);
	\fill[cyan!40!white] (2,1) rectangle (3,2);
	\fill[red!40!white] (0,2) rectangle (2,4);
	\fill[green!40!white] (2,2) rectangle (4,4);
	\draw[blue, thick] (1,1) -- (1,3);
	\node at (1.0,1.0) {$0000$};
	\node at (1.0,3.0) {$0100$};
	\node at (2.5,0.5) {$1000$};
	\node at (3.0,3.0) {$1100$};
	\node at (3.5,0.5) {$1010$};
	\node at (2.5,1.5) {$1001$};
	\node at (3.5,1.5) {$1011$};
	\draw[blue, thick] (1,1) -- (1,3);
	\draw[blue, thick] (1,3) -- (2.5,0.5);
	\draw[blue, thick] (2.5,0.5) -- (2.5,1.5);
	\draw[blue, thick] (2.5,1.5) -- (3.5,0.5);
	\draw[blue, thick] (3.5,0.5) -- (3.5,1.5);
	\draw[blue, thick] (3.5,1.5) -- (3.0,3.0);
	\end{tikzpicture}
	\caption{Morton code representation of a two-dimensional mesh with two levels of adaptation.}
	\label{fig:simple}
\end{figure}

Let us now demonstrate refinement.  At each level, the Morton code corresponds to three bits in three-dimensional space and two bits in two-dimensional space. The Morton code after refinement is generated by concatenating every bits at every level, starting from the left with the top level node in the tree and placing each node's bit representation to the right as we traverse to the bottom of the tree. A two-dimensional mesh with two levels of adaptation is presented in Fig. \ref{fig:simple} to illustrate Morton code generation with refinement. Here, we considered a two-dimensional case for ease of presentation. The final code is generated by concatenating two bits per level of adaptation. The length of the codes are extended to 4 bits by zero padding to be consistent.

An octree for a three-dimensional mesh with three levels of adaptation is shown in Fig. \ref{fig:morton1}. Here, the procedure is illustrated with coordinated colors for node 23. The Morton code for node number 23 is $001,000,101$ as shown in Fig. \ref{fig:sample}. Note that the change of color in Fig. \ref{fig:sample} coordinates with Fig. \ref{fig:morton1} to highlight placement of bits from left to right.
 
Morton code enables us to keep the elements that have closer keys allocated close to each other in the hash table. It is important to mention that the cache performance also improves due to this locality. There are several methods for Morton encoding \cite{NVIDIA, lewiner2010fast}.

\tikzset{
	stdNode/.style={rounded corners, draw, align=right},
	Myred/.style={stdNode, top color=white, bottom color=red},
	Myorange/.style={stdNode, top color=white, bottom color=orange},
	Mygreen/.style={stdNode, top color=white, bottom color=yellow}
}
\begin{figure}[h]
	\begin{tikzpicture}[sibling distance=3em, every node/.style={rounded corners, draw, align=center, top color=white, bottom color=blue!30}]] 
	\node {000 \\ (0)} 
	child { node{000 \\ (1)}}
	child { node{100 \\ (2)}} 
	child { node {010 \\ (3)}}
	child { node[Myred,align=center] {001 \\ (4)}
		child{ node[Myorange, align=center] {000 \\ (9)}
			child{ node{000\\ (17)}} 
			child { node {100\\ (18)}}
			child { node {010 \\ (19)}}
			child { node {001 \\ (20)}}
			child { node {110 \\ (21)}}
			child { node {011 \\ (22)}}
			child { node[Mygreen, align=center] {101\\ (23)}}
			child { node {111\\ (24)}} } 
		child { node {100 \\(10)} } 
		child { node {010 \\(11)} } 
		child { node {001 \\ (12) } }
		child { node {101 \\ (13)} } 
		child { node {011 \\ (14)} }
		child { node {110 \\ (15)} } 
		child { node {111 \\ (16)}} }
	child { node {101 \\ (5)}} 
	child { node {011 \\ (6)}}
	child { node {110 \\ (7) }}
	child { node {111 \\ (8) }};
	\end{tikzpicture}
	\caption{Morton code construction procedure illustrated for the node 23, The numbers in the parentheses represent the
		node index as an integer number}
	\label{fig:morton1}
\end{figure}
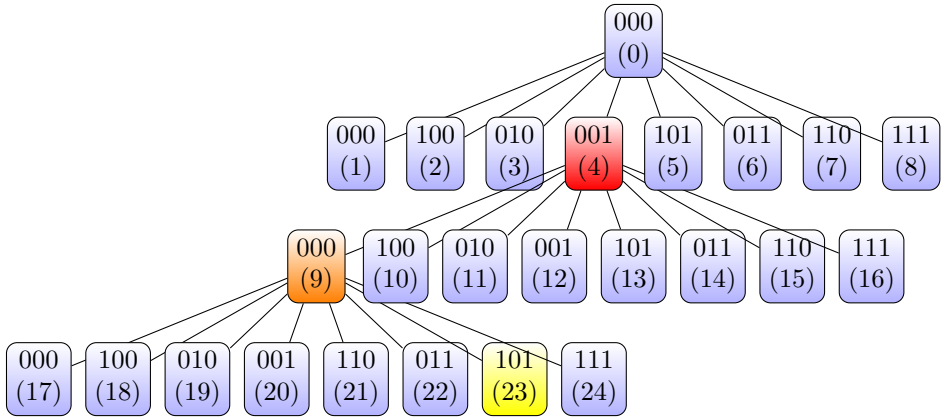

\begin{figure}[h]
	\centering
	\begin{tikzpicture}[sibling distance=3em,
	every node/.style = { rounded corners,
		draw, align=center,
		left color=red, right color=yellow!120}]]
	\node {001,000,101};
	\end{tikzpicture}
	\caption{Morton code constructed for node 23 after three levels of refinement}
	\label{fig:sample}
\end{figure}
\noindent

The procedure described above limits the number of levels for adaptation to 10 for natively 32-bit systems and to 21 for natively 64-bit systems, Our statement is valid if the AMR algorithm at hand has to convert the Morton code to an integer number because it would produce an overflow.


We should note that in the hashed linear octree approach, any hash function that does not use the entire bits and only selects some portion of the Morton code would be able to attain adaptation levels deeper than the aforementioned values at the expense of performance due to collisions in the hash table. As we have shown in \citet{hasbestan2017shortnote} Z-order curve is no longer retained once collisions occur. \citet{lewiner2010fast} discuss several approaches pertinent to the use of hash table with Morton codes.

As we discuss later in this paper, we can overcome this limitation on depth of adaptation without a performance hit by using the red-black tree data structure in the AMR algorithm. C++ \texttt{bitset} class enables defining the number of required bits beyond what is natively supported by the hardware, and the red-black tree operates on bit comparison and therefore there is no need for converting the Morton code to an integer.
 
\subsection{Hashed linear octree generation}
Data-locality is key to superior memory performance. As we mentioned earlier, one option to enhance data-locality is to store data using space-filling curves. The use of Z-order curve \cite{morton1966computer} is popular because of its easy computer implementation. Use of hash tables is also common in pointerless linear octree generation. A hash table potentially provide $O(1)$ access to its elements owing to the use of array as the underlying data-structure. However, $O(1)$ access is only guaranteed as long as there are no collisions in the hash table. In practice, this requirement is difficult to fulfill and collisions do happen when the hash function generates the same index for different keys. Typically, a remainder operator is used as the hash function to fit the keys inside a given array. 

The state-of-the-art in linear octree generation is to use a hash table with Morton codes. This choice combines the efficiency of a hash table with data-locality offered by the Z-order curve to further improve the performance of the octree generation. Morton codes are used as keys to the hash table. \cite{baert2014out, lewiner2010fast}. The works by \citet{sundar2008bottom} and \citet{burstedde2011p4est} fall under this category. 

There are several ways to construct the Morton code. In our approach, Morton codes are generated recursively from the octree hierarchy and carry all the information about the octree with them for each node of the octree. We explain the details in section \ref{AMR_infrastructure}.

In this part of our implementation, we make use of hash maps, which is a well-known associative container data structure in the field of computer science. A hash map uses a hash function to produce an index for each key. An efficient implementation of a hash map \texttt{unordered\_map} is readily available as part of the C++ standard library. C++ \texttt{unordered\_map} class, is an associative container. It associates every key to a mapped value. The mapped values are stored in buckets in such a way that the elements with key values close to each other are also stored in nearby buckets, which is good for performance. When solving a PDE, each key is mapped to a pointer to the solution vector, which is dependent on the numerical formulation and the discretization method. For example, for a cell-centered finite-volume scheme, this pointer will point to an array of unknowns in each cell, whereas in the case of a finite-element method it will point to the unknowns at the nodes of each element.  

\begin{figure}[htbp]
 	\includegraphics[scale=0.45, trim=0 7.5cm 0cm 5cm, clip]{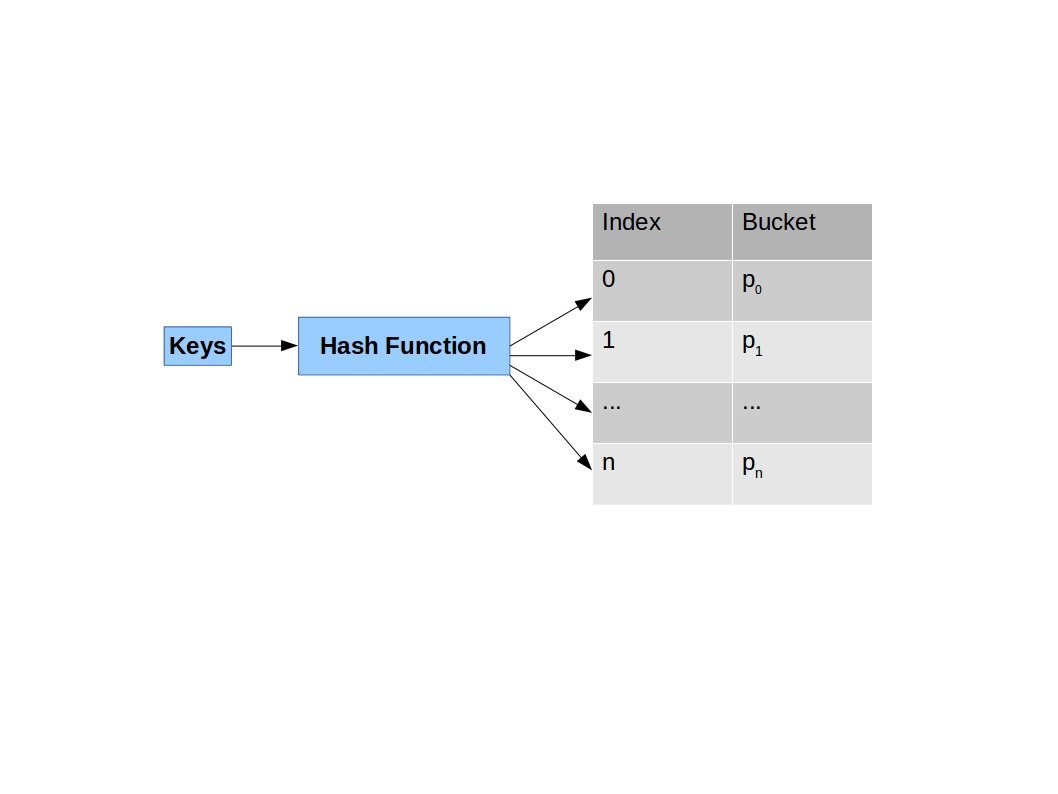}
 	\caption{Illustration of hash map data structure, each key is mapped to an index by hash function and each index corresponds to a pointer held in a bucket}
 	\label{fig:hashmap}
 \end{figure}

The hash map data structure is schematically illustrated in Fig. \ref{fig:hashmap}. (We use the term hash table to refer to the fundamental data structure, and hash map to refer to the C++ class.) The only restriction of a hash map is that the key has to be unique. The uniqueness condition is readily satisfied for the present application since every node has a unique Morton code.
The only disadvantage of the hash map associative container is that rehashing might occur if the final size of the container is not well predicted. We note that rehashing is an expensive operation computationally. If the size of the array in hash map is big enough data-locality can be preserved and perfect hashing would be possible. However, this is hard to achieve in practice for dynamic problems.
%
%
%

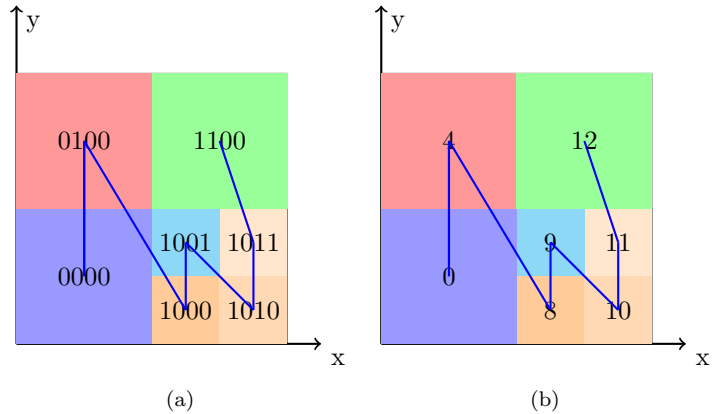
\begin{figure}[h!]
	\centering
	\subfigure[]{
		\begin{tikzpicture}[scale=0.9]
		\draw[step=2cm,gray,very thin] (0,0) grid (4,4);
		\draw[step=1cm,gray,very thin] (2,0) grid (4,2);
		\draw[thick,->] (0,0) -- (4.5,0) node[anchor=north west] {x};
		\draw[thick,->] (0,0) -- (0,5.0) node[anchor=north west] {y};;
		\fill[blue!40!white] (0,0) rectangle (2,2);
		\fill[orange!40!white] (2,0) rectangle (3,1);
		\fill[orange!30!white] (3,0) rectangle (4,1);
		\fill[orange!20!white] (3,1) rectangle (4,2);
		\fill[cyan!40!white] (2,1) rectangle (3,2);
		\fill[red!40!white] (0,2) rectangle (2,4);
		\fill[green!40!white] (2,2) rectangle (4,4);
		\draw[blue, thick] (1,1) -- (1,3);
		\node at (1.0,1.0) {$0000$};
		\node at (1.0,3.0) {$0100$};
		\node at (2.5,0.5) {$1000$};
		\node at (3.0,3.0) {$1100$};
		\node at (3.5,0.5) {$1010$};
		\node at (2.5,1.5) {$1001$};
		\node at (3.5,1.5) {$1011$};
		\draw[blue, thick] (1,1) -- (1,3);
		\draw[blue, thick] (1,3) -- (2.5,0.5);
		\draw[blue, thick] (2.5,0.5) -- (2.5,1.5);
		\draw[blue, thick] (2.5,1.5) -- (3.5,0.5);
		\draw[blue, thick] (3.5,0.5) -- (3.5,1.5);
		\draw[blue, thick] (3.5,1.5) -- (3.0,3.0);
		\end{tikzpicture}
		\label{fig:Z-curve-a}
	}
	\subfigure[]{
		\begin{tikzpicture}[scale=0.9]
		\draw[step=2cm,gray,very thin] (0,0) grid (4,4);
		\draw[step=1cm,gray,very thin] (2,0) grid (4,2);
		\draw[thick,->] (0,0) -- (4.5,0) node[anchor=north west] {x};
		\draw[thick,->] (0,0) -- (0,5.0) node[anchor=north west] {y};;
		\fill[blue!40!white] (0,0) rectangle (2,2);
		\fill[orange!40!white] (2,0) rectangle (3,1);
		\fill[orange!30!white] (3,0) rectangle (4,1);
		\fill[orange!20!white] (3,1) rectangle (4,2);
		\fill[cyan!40!white] (2,1) rectangle (3,2);
		\fill[red!40!white] (0,2) rectangle (2,4);
		\fill[green!40!white] (2,2) rectangle (4,4);
		\draw[blue, thick] (1,1) -- (1,3);
		\node at (1.0,1.0) {$0$};
		\node at (1.0,3.0) {$4$};
		\node at (2.5,0.5) {$8 $};
		\node at (3.0,3.0) {$12$};
		\node at (3.5,0.5) {$10$};
		\node at (2.5,1.5) {$9$ };
		\node at (3.5,1.5) {$11$ };
		\draw[blue, thick] (1,1) -- (1,3);
		\draw[blue, thick] (1,3) -- (2.5,0.5);
		\draw[blue, thick] (2.5,0.5) -- (2.5,1.5);
		\draw[blue, thick] (2.5,1.5) -- (3.5,0.5);
		\draw[blue, thick] (3.5,0.5) -- (3.5,1.5);
		\draw[blue, thick] (3.5,1.5) -- (3.0,3.0);
		\end{tikzpicture}
		\label{fig:Z-curve-b}
	}
	
	\caption{Illustration of mesh element indexing that arise in the hash table data structure, (a) Morton code representation of a mesh with two levels of adaptation, (b) Hash values corresponding to the Morton code from (a)}
	\label{fig:Z-curve}

\end{figure}

Theoretically, a hash table can have $O(1)$ time for insertion, removal and search operations overits elements. However, this is only possible when there are no collisions, which is known as perfect hashing. Let us explore the conditions for perfect hashing in the realm of linear octree representation. First of all, the hash function maps each key to an integer index. It is a common practice to use the remainder operator in the hash function to fit a given number of elements to an array, because a hash table uses array as its underlying data structure. Note that the operation here is limited by the maximum integer number representable on the computer architecture at hand. Consider the mesh given in Fig. \ref{fig:Z-curve-a}. In order to have a perfect hash, we first convert the Morton code to integer to produce an index and then allocate enough memory. For this simple case the size of the array should be at least 13  

To illustrate the collision issue with the hash table approach, we assume that the initial size of the hash table is 8. Now let us try to fit the 7 elements shown in Fig. \ref{fig:Z-curve-a} in the 8 slots by using the remainder operator in the hash function. For elements with hash values of 4 $(4 \bmod 8 = 4)$ and 12 $(12 \bmod 8 = 4)$ the calculated index is the same. This implies a collision will occur in the fifth slot. In other words, both elements are assigned to the same slot in the hash table. To conform to the Z-order curve and have zero collisions at the same time, the size of the array should be at least 13, which is a substantial increase in allocated memory. We also need to allocate extra space in the array for possible refinements that might occur in the next step. Obviously, this remedy to avoid collision in the hash table can be prohibitively expensive in terms of memory when a mesh with a deep-level of adaptation is desired. Subsequently, one has to relax the no-collision condition and try to use the remainder operator to fit the mesh elements in the array. Once the no-collision condition is relaxed, $O(1)$ access time is no longer guaranteed. Another draw back of using a hash table is that rehashing might occur if one does not predict the final octree size with a good approximation, and rehashing is computationally expensive. Any collision will also break the Z-order curve compliance, leading to performance loss. To have a Z-order curve with a regularly increasing index (i.e., 0, 1, 2, ...) a sorting process is inevitable, which is not readily available in a hashed linear octree generation method. To address this issue we propose to use the red-black tree in place of the hash table. 

\subsection{Binarized octree generation}
Self-balancing binary search trees, such as a red-black tree, can be favored over a hash table in certain applications, because the worst-case performance of a red-black tree for insertion, removal and search operations is guaranteed to be $O(\log N)$ . A red-black tree also keeps its elements sorted as opposed to the out-of-order storage of elements in a hash table. Red-black trees  are widely used in different areas of the computer science field \cite{guibas1978dichromatic}. Its applications ranges from computational geometry to process scheduling in Linux kernels. It guarantees that the path from the root to the farthest node is no more than twice as long as the path from the root to the nearest node \cite{drozdek2012data, schneier1992red}. In the present study, we propose to adopt the red-black tree in place of the hash table in the design of the AMR algorithm. Our goal is to create an AMR algorithm that does not lose its Z-order curve compliance at deep levels of adaptation.

A red-black tree is a self-balancing binary search tree with an extra bit for defining the color of a node. The color can be either red or black for a node. The color bit is needed to keep the tree balanced. A red-black tree obeys the following rules:
\begin{itemize}
	\item The root node is always black
	\item A red node can not have a red child
	\item Every path from a given node to any of its descendant nodes contains the same number of black nodes
\end{itemize}

The memory usage of a red-black tree is much less than an explicit octree. For a large mesh, data-locality due to strict compliance with the Z-order curve offsets the $O(\log N)$ element access cost of a red-black tree, as we have demonstrated in \cite{hasbestan2017shortnote}.

An associated container that uses a red-black tree data-structure is readily available in the C++ standard template library (STL) as \texttt{Set/Map} classes. \texttt{Set} class is a special case of the \texttt{Map} class in which the associated value is the key itself. The red-black tree needs a comparison function to insert or remove elements. We use the Morton code representation of the octree nodes as the comparison criteria for insertion of the new elements. Note that to compare two keys represented by bits, conversion of the key to an integer number is not required anymore with a red-black tree. We make use of this feature of the red-black tree in conjuction with the \texttt{bitset} class in C++ STL to generate meshes with deep levels of adaptation. We use the fast bitwise operation (xor) to perform this comparison.  

By adopting a red-black in the AMR algorithm, we accomplish the following:
\begin{enumerate}
	\item Remove the limitation on maximum level of adaptation. With a red-black tree there is no need to convert the Morton code to an integer index. Red-black tree only requires the comparison of the bits. 
	\item Preserve the order of the Z-curve exactly and ensure data-locality in the computer implementation for improved performance.
\end{enumerate}

\subsection{Graph-based unstructured mesh approach}
The Cartesian adaptive mesh generation can also be implemented using a graph-based data structure by treating the mesh as an unstructured grid topology. In this method, the connectivity for each element has to be stored. For a large mesh with a deep level of adaptation this approach requires a lot of memory because of the need to store connectivity. We have presented an unstructured approach in sufficient detail in \cite{hasbestan2017parallel} and released it as an open-source software \cite{puAMR}. Here, we will make use of it to provide a performance perspective to the other two approaches that we presented above.   

\section{AMR infrastructure methods}\label{AMR_infrastructure}
When numerically solving a PDE on an adaptively refined mesh, the AMR library needs to provide certain mesh information, such as connectivity to neighboring elements, coordinates of the elements' centroid and vertices. In this section, we explain the methods to compute such information nimbly by manipulating the Morton code representation of an octree. The methods presented in this section are common to both the hashed linear and binarized octree generation methods.

\subsection{Adaptation level}
We need to know the current adaptation level of the element under consideration as well as the neighboring elements when constructing a 2:1 balanced octree.  The Morton code representation of the octree helps a great deal to extract the level of adaptation without extra storage, because we only need to check the non-zero triple digit. 

As an example, let us again consider the node 23 in Fig. \ref{fig:sample}. The Morton code representation of node 23 is given in Fig. \ref{fig:morton1}. On a 16-bit system, Morton code would be as shown in Fig. \ref{fig:16bit}. Note that the Morton code is padded to the right with zeros to create a 16-bit system representation. Starting from the rightmost digit and scanning to the left, the last non-zero element occurs at level 3, therefore, the level of adaptation for this element is 3. Note that in the present study, element level 0 is the root element and we do not include it in the bit representation. 

\begin{figure}[h!]
	\centering
\begin{tikzpicture}[sibling distance=3em,
  every node/.style = { rounded corners,
    draw, align=center,
    top color=white, bottom color=red!30}]]
    \node {001,000,101,000,000,0}; \newline
\end{tikzpicture}
\caption{Morton code for node 23 in Fig. \ref{fig:morton1} on a prototypical 16-bit system.}
\label{fig:16bit}
\end{figure}
\noindent

For the special case of an element represented with all zeros as Morton code, we need to employ a trick. This special case of all zero bit element becomes important while enforcing the 2:1 balance. Because all digits are zeros, we cannot extract the adaptation level, but we can manipulate the siblings to extract this information. Note that the maximum \texttt{bitset} instance size is preset before the refinement as this is a template class where the size of the object is the template parameter. One can start by flipping any one of the digits belonging to the maximum size. Note that by flipping the bit we are now operating on the sibling now. After this we check for the existence of such a key. If it exists then the element under consideration and the sibling belong to the same level.  

\subsection{Edge length}
Edge length is required to calculate volume and surface integrals. Once the adaptation level is known following the above procedure, it is straightforward to calculate the edge length resulting from an isotropic refinement in  three-dimensional space, which is simply a generalization of one-dimensional
bisection procedure. 

Given the edge length at the top level ${l_{0}}$, the length of the segment at each adaptation level ${i}$ can be calculated as follows: 

\begin{align}
l_{i}=\frac{l_{0}}{2^{i}}, i=1, 2, ...,nLevel.
\end{align}
\subsection{Centroid coordinates}
In the solution of fluid flow equations, whether a collocated or staggered arrangement of the variables is pursued, the centroid of the element needs to be known. Given the initial element's centroid and length of the edges of the initial element in $x, y$ and $z$ directions, it is possible to calculate the centroid of the 
current element as follows:

\begin{align}
x_{i}=x_{i}^{0}-({(-1)^{bit[3 j+i]}} \frac{1}{2^{j+2}}) l_{i}, \\ \nonumber
j=0, 1, ... , nLevel-1, \\
i=1,2,3. \nonumber
\end{align}
\noindent
where $x_{i}^{0}$ implies the $x_{i}$ component of the centroid of the initial element. Note that summation is implied on the ${j}$ index. 
To illustrate with an example, the coordinates of the centroid for the Morton code given in Fig. \ref{fig:sample} are calculated as follows:

\begin{align}
x_{c}=x_{0}+l_{0}({-\frac{1}{4}-\frac{1}{8}+\frac{1}{16}}), \nonumber\\
y_{c}=y_{0}+l_{0}({-\frac{1}{4}-\frac{1}{8}-\frac{1}{16}}), \\
z_{c}=z_{0}+l_{0}({+\frac{1}{4}-\frac{1}{8}+\frac{1}{16}}).\nonumber
\end{align}

\subsection{Element vertices}
With an element's centroid and edge length at hand, the vertices of each element
is obtained by adding and subtracting the half of the edge length to the centroid at each direction as follows:

\begin{align}
x=x_{c}\pm 0.5 \times l_{0}, \nonumber\\
y=y_{c}\pm 0.5 \times l_{0}, \\
z=z_{c}\pm 0.5 \times l_{0}. \nonumber
\end{align}

\subsection{Connectivity to neighbors}
Connectivity to surrounding elements is essential information when solving PDEs, whether it is for calculating the flux of a quantity for enforcing conservation principles or constructing the Jacobian (or stiffness) matrices for implicit schemes. The way connectivity information is stored and accessed in computer memory affects the overall performance tremendously. In this section, we discuss the construction of connectivity that is unique to our implementation. We consider a two-dimensional example for ease of illustration, but our implementation is for the three-dimensional case. Before proceeding further, we define our terminology.

\subsubsection{Siblings}
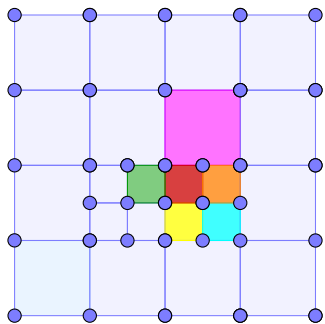
\begin{figure}[h]
	\centering
	\pagestyle{empty}
	\definecolor{qqzzqq}{rgb}{0.,0.6,0.}
	\definecolor{ffxfqq}{rgb}{1.,0.4980392156862745,0.}
	\definecolor{ccqqqq}{rgb}{0.8,0.,0.}
	\definecolor{qqffff}{rgb}{0.,1.,1.}
	\definecolor{ffffqq}{rgb}{1.,1.,0.}
	\definecolor{ffqqff}{rgb}{1.,0.,1.}
	\definecolor{zzccff}{rgb}{0.6,0.8,1.}
	\definecolor{xdxdff}{rgb}{0.49019607843137253,0.49019607843137253,1.}
	\begin{tikzpicture}[line cap=round,line join=round,>=triangle 45,x=1.0cm,y=1.0cm]
	\fill[color=zzccff,fill=zzccff,fill opacity=0.20000000298023224] (0.,0.) -- (1.,0.) -- (1.,1.) -- (0.,1.) -- cycle;
	\fill[color=xdxdff,fill=xdxdff,fill opacity=0.10000000149011612] (1.,0.) -- (2.,0.) -- (2.,1.) -- (1.,1.) -- cycle;
	\fill[color=xdxdff,fill=xdxdff,fill opacity=0.10000000149011612] (3.,1.) -- (2.,1.) -- (2.,0.) -- (3.,0.) -- cycle;
	\fill[color=xdxdff,fill=xdxdff,fill opacity=0.10000000149011612] (4.,1.) -- (3.,1.) -- (3.,0.) -- (4.,0.) -- cycle;
	\fill[color=xdxdff,fill=xdxdff,fill opacity=0.10000000149011612] (1.,1.) -- (1.5,1.) -- (1.5,1.5) -- (1.,1.5) -- cycle;
	\fill[color=xdxdff,fill=xdxdff,fill opacity=0.10000000149011612] (2.,1.) -- (2.,1.5) -- (1.5,1.5) -- (1.5,1.) -- cycle;
	\fill[color=xdxdff,fill=xdxdff,fill opacity=0.10000000149011612] (0.,1.) -- (1.,1.) -- (1.,2.) -- (0.,2.) -- cycle;
	\fill[color=xdxdff,fill=xdxdff,fill opacity=0.10000000149011612] (0.,2.) -- (1.,2.) -- (1.,3.) -- (0.,3.) -- cycle;
	\fill[color=xdxdff,fill=xdxdff,fill opacity=0.10000000149011612] (1.,2.) -- (2.,2.) -- (2.,3.) -- (1.,3.) -- cycle;
	\fill[color=xdxdff,fill=xdxdff,fill opacity=0.10000000149011612] (0.,3.) -- (1.,3.) -- (1.,4.) -- (0.,4.) -- cycle;
	\fill[color=xdxdff,fill=xdxdff,fill opacity=0.10000000149011612] (1.,3.) -- (2.,3.) -- (2.,4.) -- (1.,4.) -- cycle;
	\fill[color=xdxdff,fill=xdxdff,fill opacity=0.10000000149011612] (2.,3.) -- (3.,3.) -- (3.,4.) -- (2.,4.) -- cycle;
	\fill[color=xdxdff,fill=xdxdff,fill opacity=0.10000000149011612] (3.,3.) -- (4.,3.) -- (4.,4.) -- (3.,4.) -- cycle;
	\fill[color=ffqqff,fill=ffqqff,fill opacity=0.550000011920929] (2.,2.) -- (3.,2.) -- (3.,3.) -- (2.,3.) -- cycle;
	\fill[color=xdxdff,fill=xdxdff,fill opacity=0.10000000149011612] (3.,2.) -- (4.,2.) -- (4.,3.) -- (3.,3.) -- cycle;
	\fill[color=xdxdff,fill=xdxdff,fill opacity=0.10000000149011612] (3.,1.) -- (4.,1.) -- (4.,2.) -- (3.,2.) -- cycle;
	\fill[color=ffffqq,fill=ffffqq,fill opacity=0.75] (2.,1.) -- (2.5,1.) -- (2.5,1.5) -- (2.,1.5) -- cycle;
	\fill[color=qqffff,fill=qqffff,fill opacity=0.75] (2.5,1.) -- (3.,1.) -- (3.,1.5) -- (2.5,1.5) -- cycle;
	\fill[color=ccqqqq,fill=ccqqqq,fill opacity=0.75] (2.,1.5) -- (2.5,1.5) -- (2.5,2.) -- (2.,2.) -- cycle;
	\fill[color=ffxfqq,fill=ffxfqq,fill opacity=0.75] (2.5,1.5) -- (3.,1.5) -- (3.,2.) -- (2.5,2.) -- cycle;
	\fill[color=xdxdff,fill=xdxdff,fill opacity=0.10000000149011612] (1.,1.5) -- (1.5,1.5) -- (1.5,2.) -- (1.,2.) -- cycle;
	\fill[color=qqzzqq,fill=qqzzqq,fill opacity=0.5] (1.5,1.5) -- (2.,1.5) -- (2.,2.) -- (1.5,2.) -- cycle;
	\draw [color=xdxdff] (0.,0.)-- (1.,0.);
	\draw [color=zzccff] (1.,0.)-- (1.,1.);
	\draw [color=zzccff] (1.,1.)-- (0.,1.);
	\draw [color=xdxdff] (0.,1.)-- (0.,0.);
	\draw [color=xdxdff] (1.,0.)-- (2.,0.);
	\draw [color=xdxdff] (2.,0.)-- (2.,1.);
	\draw [color=xdxdff] (2.,1.)-- (1.,1.);
	\draw [color=xdxdff] (1.,1.)-- (1.,0.);
	\draw [color=xdxdff] (3.,1.)-- (2.,1.);
	\draw [color=xdxdff] (2.,1.)-- (2.,0.);
	\draw [color=xdxdff] (2.,0.)-- (3.,0.);
	\draw [color=xdxdff] (3.,0.)-- (3.,1.);
	\draw [color=xdxdff] (4.,1.)-- (3.,1.);
	\draw [color=xdxdff] (3.,1.)-- (3.,0.);
	\draw [color=xdxdff] (3.,0.)-- (4.,0.);
	\draw [color=xdxdff] (4.,0.)-- (4.,1.);
	\draw [color=xdxdff] (1.,1.)-- (1.5,1.);
	\draw [color=xdxdff] (1.5,1.)-- (1.5,1.5);
	\draw [color=xdxdff] (1.5,1.5)-- (1.,1.5);
	\draw [color=xdxdff] (1.,1.5)-- (1.,1.);
	\draw [color=xdxdff] (2.,1.)-- (2.,1.5);
	\draw [color=xdxdff] (2.,1.5)-- (1.5,1.5);
	\draw [color=xdxdff] (1.5,1.5)-- (1.5,1.);
	\draw [color=xdxdff] (1.5,1.)-- (2.,1.);
	\draw [color=xdxdff] (0.,1.)-- (1.,1.);
	\draw [color=xdxdff] (1.,1.)-- (1.,2.);
	\draw [color=xdxdff] (1.,2.)-- (0.,2.);
	\draw [color=xdxdff] (0.,2.)-- (0.,1.);
	\draw [color=xdxdff] (0.,2.)-- (1.,2.);
	\draw [color=xdxdff] (1.,2.)-- (1.,3.);
	\draw [color=xdxdff] (1.,3.)-- (0.,3.);
	\draw [color=xdxdff] (0.,3.)-- (0.,2.);
	\draw [color=xdxdff] (1.,2.)-- (2.,2.);
	\draw [color=xdxdff] (2.,2.)-- (2.,3.);
	\draw [color=xdxdff] (2.,3.)-- (1.,3.);
	\draw [color=xdxdff] (1.,3.)-- (1.,2.);
	\draw [color=xdxdff] (0.,3.)-- (1.,3.);
	\draw [color=xdxdff] (1.,3.)-- (1.,4.);
	\draw [color=xdxdff] (1.,4.)-- (0.,4.);
	\draw [color=xdxdff] (0.,4.)-- (0.,3.);
	\draw [color=xdxdff] (1.,3.)-- (2.,3.);
	\draw [color=xdxdff] (2.,3.)-- (2.,4.);
	\draw [color=xdxdff] (2.,4.)-- (1.,4.);
	\draw [color=xdxdff] (1.,4.)-- (1.,3.);
	\draw [color=xdxdff] (2.,3.)-- (3.,3.);
	\draw [color=xdxdff] (3.,3.)-- (3.,4.);
	\draw [color=xdxdff] (3.,4.)-- (2.,4.);
	\draw [color=xdxdff] (2.,4.)-- (2.,3.);
	\draw [color=xdxdff] (3.,3.)-- (4.,3.);
	\draw [color=xdxdff] (4.,3.)-- (4.,4.);
	\draw [color=xdxdff] (4.,4.)-- (3.,4.);
	\draw [color=xdxdff] (3.,4.)-- (3.,3.);
	\draw [color=ffqqff] (2.,2.)-- (3.,2.);
	\draw [color=ffqqff] (3.,2.)-- (3.,3.);
	\draw [color=ffqqff] (3.,3.)-- (2.,3.);
	\draw [color=ffqqff] (2.,3.)-- (2.,2.);
	\draw [color=xdxdff] (3.,2.)-- (4.,2.);
	\draw [color=xdxdff] (4.,2.)-- (4.,3.);
	\draw [color=xdxdff] (4.,3.)-- (3.,3.);
	\draw [color=xdxdff] (3.,3.)-- (3.,2.);
	\draw [color=xdxdff] (3.,1.)-- (4.,1.);
	\draw [color=xdxdff] (4.,1.)-- (4.,2.);
	\draw [color=xdxdff] (4.,2.)-- (3.,2.);
	\draw [color=xdxdff] (3.,2.)-- (3.,1.);
	\draw [color=ffffqq] (2.,1.)-- (2.5,1.);
	\draw [color=ffffqq] (2.5,1.)-- (2.5,1.5);
	\draw [color=ffffqq] (2.5,1.5)-- (2.,1.5);
	\draw [color=ffffqq] (2.,1.5)-- (2.,1.);
	\draw [color=qqffff] (2.5,1.)-- (3.,1.);
	\draw [color=qqffff] (3.,1.)-- (3.,1.5);
	\draw [color=qqffff] (3.,1.5)-- (2.5,1.5);
	\draw [color=qqffff] (2.5,1.5)-- (2.5,1.);
	\draw [color=ccqqqq] (2.,1.5)-- (2.5,1.5);
	\draw [color=ccqqqq] (2.5,1.5)-- (2.5,2.);
	\draw [color=ccqqqq] (2.5,2.)-- (2.,2.);
	\draw [color=ccqqqq] (2.,2.)-- (2.,1.5);
	\draw [color=ffxfqq] (2.5,1.5)-- (3.,1.5);
	\draw [color=ffxfqq] (3.,1.5)-- (3.,2.);
	\draw [color=ffxfqq] (3.,2.)-- (2.5,2.);
	\draw [color=ffxfqq] (2.5,2.)-- (2.5,1.5);
	\draw [color=xdxdff] (1.,1.5)-- (1.5,1.5);
	\draw [color=xdxdff] (1.5,1.5)-- (1.5,2.);
	\draw [color=xdxdff] (1.5,2.)-- (1.,2.);
	\draw [color=xdxdff] (1.,2.)-- (1.,1.5);
	\draw [color=qqzzqq] (1.5,1.5)-- (2.,1.5);
	\draw [color=qqzzqq] (2.,1.5)-- (2.,2.);
	\draw [color=qqzzqq] (2.,2.)-- (1.5,2.);
	\draw [color=qqzzqq] (1.5,2.)-- (1.5,1.5);
	\begin{scriptsize}
	\draw [fill=xdxdff] (0.,0.) circle (1.5pt);
	\draw [fill=xdxdff] (1.,0.) circle (2.5pt);
	\draw [fill=xdxdff] (1.,1.) circle (2.5pt);
	\draw [fill=xdxdff] (1.,1.) circle (1.5pt);
	\draw [fill=xdxdff] (2.,1.) circle (2.5pt);
	\draw [fill=xdxdff] (2.,0.) circle (2.5pt);
	\draw [fill=xdxdff] (2.,1.) circle (2.5pt);
	\draw [fill=xdxdff] (1.,1.) circle (2.5pt);
	\draw [fill=xdxdff] (3.,0.) circle (2.5pt);
	\draw [fill=xdxdff] (3.,1.) circle (2.5pt);
	\draw [fill=xdxdff] (2.,0.) circle (2.5pt);
	\draw [fill=xdxdff] (3.,0.) circle (2.5pt);
	\draw [fill=xdxdff] (4.,0.) circle (2.5pt);
	\draw [fill=xdxdff] (4.,1.) circle (2.5pt);
	\draw [fill=xdxdff] (3.,0.) circle (2.5pt);
	\draw [fill=xdxdff] (4.,0.) circle (2.5pt);
	\draw [fill=xdxdff] (1.5,1.) circle (2.5pt);
	\draw [fill=xdxdff] (1.5,1.5) circle (2.5pt);
	\draw [fill=xdxdff] (1.,1.5) circle (2.5pt);
	\draw [fill=xdxdff] (2.,1.5) circle (2.5pt);
	\draw [fill=xdxdff] (1.,2.) circle (2.5pt);
	\draw [fill=xdxdff] (0.,2.) circle (2.5pt);E
	\draw [fill=xdxdff] (1.,3.) circle (2.5pt);
	\draw [fill=xdxdff] (0.,3.) circle (2.5pt);
	\draw [fill=xdxdff] (2.,2.) circle (2.5pt);
	\draw [fill=xdxdff] (2.,3.) circle (2.5pt);
	\draw [fill=xdxdff] (1.,3.) circle (2.5pt);
	\draw [fill=xdxdff] (1.,4.) circle (2.5pt);
	\draw [fill=xdxdff] (0.,4.) circle (2.5pt);
	\draw [fill=xdxdff] (2.,4.) circle (2.5pt);
	\draw [fill=xdxdff] (1.,4.) circle (2.5pt);
	\draw [fill=xdxdff] (3.,3.) circle (2.5pt);
	\draw [fill=xdxdff] (3.,4.) circle (2.5pt);
	\draw [fill=xdxdff] (2.,4.) circle (2.5pt);
	\draw [fill=xdxdff] (4.,3.) circle (2.5pt);
	\draw [fill=xdxdff] (4.,4.) circle (2.5pt);
	\draw [fill=xdxdff] (3.,4.) circle (2.5pt);
	\draw [fill=xdxdff] (3.,2.) circle (2.5pt);
	\draw [fill=xdxdff] (3.,3.) circle (2.5pt);
	\draw [fill=xdxdff] (2.,3.) circle (2.5pt);
	\draw [fill=xdxdff] (4.,2.) circle (2.5pt);
	\draw [fill=xdxdff] (4.,3.) circle (2.5pt);
	\draw [fill=xdxdff] (3.,3.) circle (2.5pt);
	\draw [fill=xdxdff] (4.,2.) circle (2.5pt);
	\draw [fill=xdxdff] (3.,2.) circle (2.5pt);
	\draw [fill=xdxdff] (2.5,1.) circle (2.5pt);E
	\draw [fill=xdxdff] (2.5,1.5) circle (2.5pt);
	\draw [fill=xdxdff] (2.,1.5) circle (2.5pt);
	\draw [fill=xdxdff] (3.,1.5) circle (2.5pt);
	\draw [fill=xdxdff] (2.5,1.5) circle (2.5pt);
	\draw [fill=xdxdff] (2.5,2.) circle (2.5pt);
	\draw [fill=xdxdff] (2.,2.) circle (2.5pt);
	\draw [fill=xdxdff] (3.,2.) circle (2.5pt);
	\draw [fill=xdxdff] (2.5,2.) circle (2.5pt);
	\draw [fill=xdxdff] (1.5,2.) circle (2.5pt);
	\draw [fill=xdxdff] (1.,2.) circle (2.5pt);
	\draw [fill=xdxdff] (2.,2.) circle (2.5pt);
	\draw [fill=xdxdff] (1.5,2.) circle (2.5pt);
	\draw [fill=xdxdff] (-9.180327762539031E-4,1.001500354284914) circle (2.5pt);
	\draw [fill=xdxdff] (0.,0.) circle (2.5pt);
	\end{scriptsize}
	\end{tikzpicture}
	\caption{Illustration of neighbor identification from Morton code.
		Morton codes for different colors are: (00,11,11) for green; (10,01,01) for red; (01,10,00) for yellow;  (10,01,11) for orange;  (11,00) for magenta; (10,00,10) for cyan.}
	\label{fig:nbr_id}
\end{figure}

Siblings of each element, can be found by flipping the corresponding bit in $x, y, z$  directions, respectively. In this operation, only one bit is flipped. Flipping the bit in this context implies the modification of the bit to \textit{true} if it is \textit{false}, and to \textit{false} when it is originally set to \textit{true}. An example is shown in Fig. \ref{fig:nbr_id}. For the element highlighted in red, the Morton code in two-dimensional space, using the procedure explained in previous sections, can be constructed as (10,01,01). The siblings for this element can be identified by flipping the corresponding bit in $x$ and $y$ directions. Therefore, the Morton code for the element located in the right hand side is (10,01,11) and for the element underneath is  (10,01,00). The siblings are depicted in orange and yellow in Fig. \ref{fig:nbr_id}, respectively. The element highlighted in cyan is a non-neighbor sibling.

\subsubsection{Off-branch neighbors}
\label{sec:offbranch}
The method we presented in section \ref{subsection:morton} for storing an octree as Morton codes enables the identification of off-branch neighbors by only manipulating the bits. Here, the word \textit{off-branch} refers to the neighbors that are not siblings to a given element. Finding non-local siblings in each direction, in this case the green and purple element in Fig. \ref{fig:nbr_id}, consists of two steps. In the first step, we need to detect the level at which the bit was flipped at the corresponding direction. The technique to accomplish this task is presented in Algorithm \ref{alg:bitflip}.
\begin{algorithm}[h]
	\caption{Detection of bit-flip level}
	\label{alg:bitflip}
	\begin{algorithmic}
		\State {define $T$ : set of octree Morton codes}
		\State{given  $M\in T$}
		\State{define $k={1,2,3}$ : for (x, y, or z) directions, respectively}
		\State{define $d$ :  dimension of the computational space}
		\State{define $f$ : flip-level}
		\State{define $M_l$ : level of $M$}
		\State{Initialize : bolean=M(d*l+k)}
		\For{ $i=M_l-1: 0$}		
		\If{bolean!=M(d*i+k)}
		\State{$f=i$}
		\State{\textbf{break;}}
		\EndIf
		\EndFor		
		\State{Return $f$}
	\end{algorithmic}
\end{algorithm}

Let us describe it with an example. Consider the element highlighted in red in Fig. \ref{fig:nbr_id}. Morton code for this element can easily be calculated as (10,01,01). Our goal is to construct the Morton code depicted in green without an explicit search. This requires displacement in the $x-$direction (i.e., $k=1$ in \ref{alg:bitflip}). 
There are two steps involved in this task:

\begin{itemize}
\item The location of the red element in Fig. \ref{fig:nbr_id} at the finest level is $01$, so the bit in $x-$direction is $0$ (or false). Moving up one level at the tree towards parents, the location of this element is found as $01$, which corresponds to false as well. Proceeding one level further, the bit corresponding to that location becomes $1$ (true). So this level is the bit-flip level. Therefore,  moving from left to right, the bits corresponding to the $x-$direction is $(1,0,0)$. So the level at which flip has occurred goes all the way up to the first level.
\item The second step is to use the information embedded in the bit representation such that the code for the off-branch neighbor is directly extracted. To obtain the Morton code for off-branch neighbor (i.e., the green element in Fig. \ref{fig:nbr_id}) in $x-$direction, we flip all the bits in the $x-$direction up to the bit-flip level. This operation yields the Morton code as (00,10,11).
\end{itemize}

Note that by following the above procedure, there is no need for an explicit search to find the neighboring elements, because the Morton code for each element carries with itself those information in a compact fashion.

Next, we explain two special cases in which the off-branch neighbor has a different level of adaptation. Because we work with a balanced tree, such special cases can easily be resolved.

\tikzset{
	heap/.style={
		every node/.style={rounded corners, draw, align=Eright, top color=white, bottom color=blue!30}},
		level 1/.style={sibling distance=30mm},
		level 2/.style={sibling distance=7mm}
	}

\tikzset{
	stdNode/.style={rounded corners, draw, align=right},
	l0/.style={stdNode,sibling distance=30mm, top color=white, bottom color=blue!30},
	lr/.style={stdNode,sibling distance=30mm, top color=white, bottom color=red},
    lg/.style={stdNode,sibling distance=30mm, top color=white, bottom color=green},
    lm/.style={stdNode,sibling distance=30mm, top color=white, bottom color=magenta},
    ly/.style={stdNode,sibling distance=30mm, top color=white, bottom color=yellow},
    lo/.style={stdNode,sibling distance=30mm, top color=white, bottom color=orange},
    lb/.style={stdNode,sibling distance=30mm, top color=white, bottom color=cyan},
}

\begin{figure}[h]
\begin{tikzpicture} 
\node[l0] {00} 
child { node[l0]{00} 
	child{ node[l0]{00}} 
	child { node[l0] {01}}
	child { node[l0] {10}}
	child { node[l0] {11}
		child{ node[l0]{00}} 
		child { node[l0] {01}}
		child { node[l0] {10}}
		child { node[lg] {11}}
    }}
child{node[l0]{01}
	child{ node[l0]{00}} 
	child { node[l0] {01}}
	child { node[l0] {10}}
	child { node[l0] {11}}
      } 
child { node[l0] {10}
	child{ node[l0]{00} } 
	child { node[l0] {01}
		child{node[ly] {00}} 
	    child{node[lr] {01}} 
	    child{node[lb] {10}} 
	    child{node[lo] {11} }}
	child { node[l0] {10} 
	}
	child { node[l0] {11} }
     }
child { node[l0]{11}
		child{ node[lm]{00}} 
		child { node[l0] {01}}
		child { node[l0] {10}}
		child { node[l0] {11}}
        };	

\end{tikzpicture}
\caption{Representation of the mesh given in Fig. \ref{fig:nbr_id} as a tree. Matching colors are used in both figures.}
\end{figure}

\subsubsection{Case I}
The off-branch neighbor has an adaptation level lower than the element under consideration. This is the case for the element highlighted in red in Fig. \ref{fig:nbr_id}. The goal is to construct the Morton code for the element highlighted in magenta. Again all that needs to be done is to eliminate the last two bits so that the level is reduced by one. 
This will lead to Morton code of (11,00) for the magenta element in Fig. \ref{fig:nbr_id}.
Note that the length of each code is consistent with the level of refinement. 
\subsubsection{Case II}
The off-branch neighbor has an adaptation level higher than the element under consideration. To illustrate this case,  consider the element highlighted in magenta. The goal here is to go in the reverse direction and construct the Morton code of the off-branch neighbors in the $y$-direction (i.e., the elements highlighted in red and orange). This adds another step to what has already been done earlier, 
\begin{itemize}
\item The first step involves the prior two steps given in section \ref{sec:offbranch}. The Morton code for the magenta element is given as (11,00), the flip has occurred at level 2, therefore flipping the bits of this code will result in (10,01), 
\item Since the level of this element is higher than the element under consideration, it is required to repeat the last two bits, this will give (10,01,01), this element is one of the neighbors.  To construct the second neighbor all we have to do is find the sibling of this code in $x$-direction, which is (10,01,11). 
\end{itemize}
In three-dimensional space, two more neighbors are added to the list. These neighbors are constructed by alternating the bit in $x-$ and $z-$ directions.

\subsection{Detecting boundary elements}
Boundary elements are those elements that only include siblings as part of their neighborhood connectivity. Boundary elements do not possess any off-branch neighbors. We make use of bitwise operations to identify boundary elements for a given Morton code. The following theorem can be used to detect boundary elements from their Morton code representation.

\begin{theorem}
An element is a boundary element at a given direction if and only if its Morton code is free of bit-flips in that direction. 
\end{theorem}

Elements with Morton codes (10,11), (10,10) do not have any off-branch neighbors in the $x$-direction. Hence, they are boundary elements. If we traverse in the $y$-direction, elements with codes (10,00) and (10,10) are also boundary elements. The advantage of our approach is that no bits are used for tagging the boundary elements since this criteria can efficiently be extracted from the Morton code at hand. 

\section{Balancing the octree}

During the process to generate an octree with dynamic adaptation, the 2:1 balance may break. Such imbalance propagates, affecting immediate neighbors and neighbors of neighbors. This issue has been coined as the \textit{ripple effect} in the literature of balanced octree generation \cite{sundar2008bottom}.
Our computer implementation led to the algorithm described in \ref{alg:balancetree}.
It starts with an initial given queue, and extends that queue to maintain the balance for the entire tree. 

\begin{algorithm}
	\caption{ 2:1 balanced refinement-list generation}
	\label{alg:balancetree}
	\begin{algorithmic}
		\State{define $P=1$}
		\State{define $Q$ : List of elements tagged to be refined}
		\State{define $NB$ : Neighbor}
		\State{define $NB_{l}$ : Level of neighbor}
		\State{define $C_{l}$ : Level of C element}
		\State{define $istart=0$; $iend=Q.size()$}
		\While{$P > 0$}
		\State{define $P=0$}
		\For{ $i=istart: iend$}		
		\State{$C={Q(i)}$}
		\If{$NB_l < C_{l}$ $\And$ $NB\notin Q$}
		\State{append $NB$ to $Q$}
		\State{set $P=1$}		
		\EndIf
		\EndFor		
		\State{$istart= iend$; $iend = Q.size()$}
		\EndWhile
		\State{Return $Q$}
	\end{algorithmic}
\end{algorithm}

There are several ways to implement the $NB\notin Q$ section of the code.
We refer the reader to Algorithm \ref{alg:balancetree} for definitions of $NB$ and $Q$. 
The $NB\notin Q$ operation can be prohibitively expensive if not handled in a proper way. In the process of balancing, if any of the neighbors of the element under consideration violates the condition, before adding this element to the end of queue, one needs to check if that element is already in the refinement list.
When using array data structure or the \texttt{Vector} class, the cost of this operation can be calculated to grow as $O(N*N)$, where $N$ is the number of elements tagged for refinement. The cost is prohibitively large for a mesh on the order of several hundred million elements. To alleviate this problem and reduce the $O(N*N)$ cost to $m*O(N)$, where $m$ is the number of iterations done by in the while loop due to the ripple effect, we use a hash map for this list. We note here that for both the hashed linear octree and the binarized octree generation methods, a hash table data structure is used to store the list of the elements tagged for refinement. We achieve significant reduction in time spent for the balance algorithm with this modification. Timings for the Stanford bunny geometry using the \texttt{Vector} and \texttt{Map} in search are presented in Table \ref{tab:table3}. Once again, this example illustrates the value of selecting the right data structure in scientific computing.  

\begin{table}[h!]
	\centering
	\caption{Comparison of total runtime for two different data structures and their effect on the search algorithms in the 2:1 balance, the Stanford bunny geometry is used in this analysis}
	\label{tab:table3}
	\begin{tabular}{cccccc}
		\toprule
		level & 9 & 10 & 11 & 12 & 13\\
		\midrule
		mesh size & 118,357 & 437,557 & 1,199,290 & 2,830,815 & 5,867,800\\
		\midrule
		time [s](Hash Map) & 0.8 & 2.57 & 4.71 & 10.47 & 26.0\\    
		\midrule
		time [s](Vector) & 1.35 & 5.40 & 19.02 & 63.0 & 185.37\\ 
		\bottomrule
	\end{tabular}
\end{table}

\section{Tagging elements for refinement}
The mesh generation and adaptation process requires the identification of the elements which intersects the immersed geometry. The brute-force approach to obtain the list of tagged elements for refinement would be through a search. However, a brute-force approach can become expensive for a large mesh. 
With a brute-force search, the computational cost grows with $m \times n$, where $n$ is the number of points of the geometry and $m$ number of mesh elements at the current level. and ensure data-locality in the computer implementation 

One way to alleviate this bottleneck and reduce the cost of search is to voxelize the geometry and perform the search on the specific voxel instead of the entire geometry. This method improves search proportional to number of voxels. Alternatively, by taking advantage of the Morton code representation, it is possible to completely remove this search stage using \textit{geometry encoding}. Removing the search here implies dropping the cost from $O(m\times n)$ to $O(1)$ if hash table used as an underlying container for the list of tagged elements. 

Coordinates of the centroid of the STL triangles are used to determine if an element resides in a given voxel. 
Normals of the surfaces can also be assigned in a similar fashion.


\subsection{Voxelization of immersed geometry}
Voxelization of the immersed geometry comes handy to restrict certain numerical operations around the geometry. One could consider voxelization as a refined bounding box wrapping around the geometry. The map values are pointers to the geometry file portion that is segmented by the given element. We store the solid geometry in STereoLithography (STL) file format. 

The level of refinement in voxelization is much lower than the solution mesh. It is specified independent of the solution according to the solid geometry at hand. The voxel class inherits from the base tree class as shown in Listing 1. We note  that voxel need not be a 2:1 balanced octree.

\begin{figure}[h!]
	\begin{center}
		\includegraphics[width=0.6\textwidth]{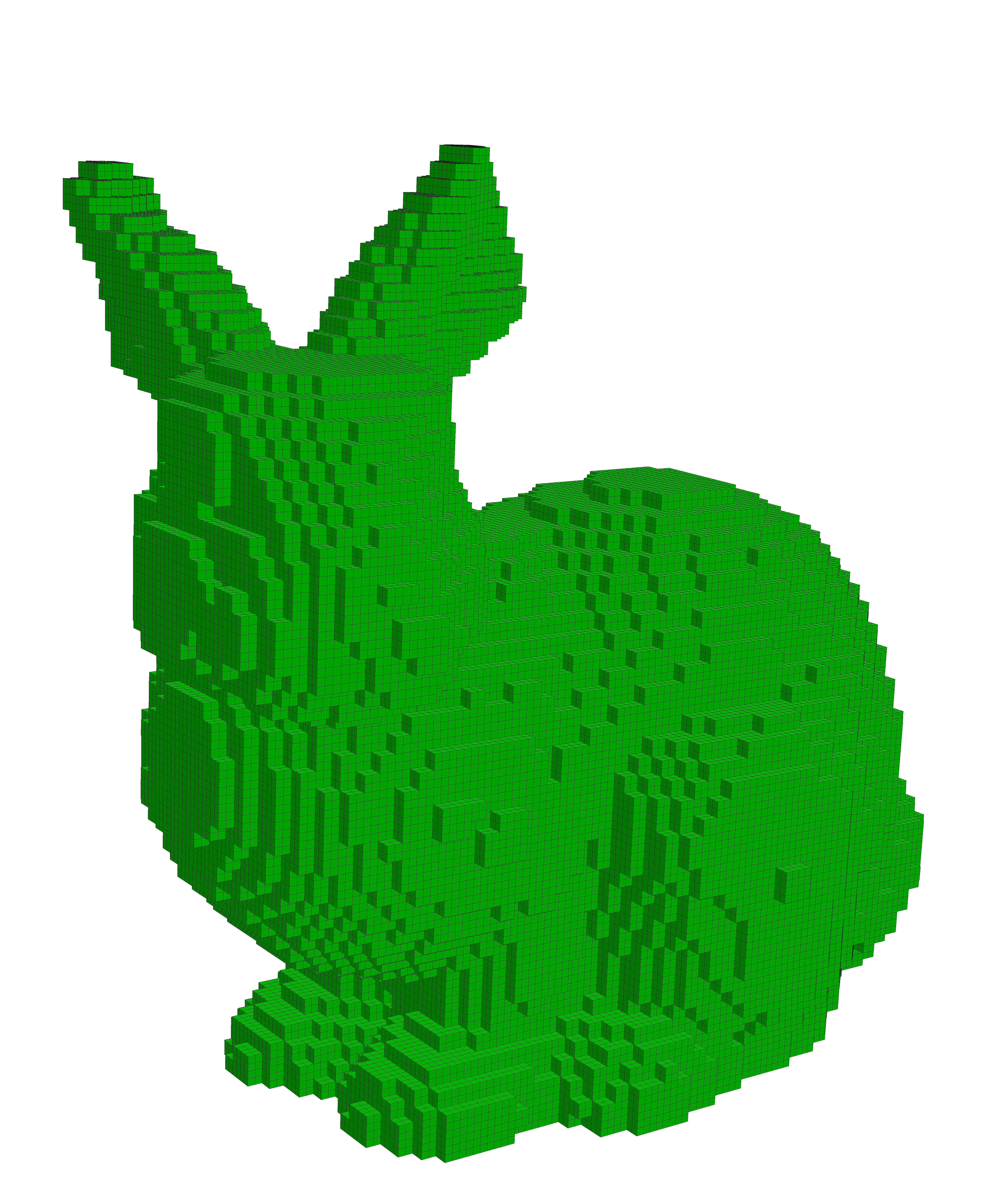}
	\end{center}
	\caption{Voxelized representation of the Stanford bunny geometry.}
	\label{fig:voxel}
\end{figure}

Voxelized view of the Stanford bunny is presented in Fig. \ref{fig:voxel}. One does not need this many voxels to accelerate the search. A higher refinement level is adopted here for illustration purposes.

\definecolor{mygreen}{rgb}{0,0.6,0}
\definecolor{mygray}{rgb}{0.5,0.5,0.5}
\definecolor{mymauve}{rgb}{0.58,0,0.82}

\lstdefinestyle{customc}{
  belowcaptionskip=1\baselineskip,
  breaklines=true,
  frame=L,
  xleftmargin=\parindent,
  language=C++,
  showstringspaces=false,
  basicstyle=\footnotesize\ttfamily,
  keywordstyle=\bfseries\color{red},
  commentstyle=\itshape\color{black},
  identifierstyle=\color{blue},
  stringstyle=\color{orange},
}
\lstset{style=customc}

\lstset{caption={Base tree class and inherited voxel class},label=class}

\lstset{basicstyle=\scriptsize}
\begin{minipage}{\linewidth}
\begin{lstlisting}[frame=single]
template <size_t N, typename value>
class Tree
{
    template <size_t N1, typename value>
    friend class Hdf5Xmf;     //write out in hdf5 format

    protected:
    real      rootlength[3]; // length of the first generation
    real      rootcoords[3]; // centeroid of the first generation
    bitset<N> rootkey = 0;   // root value is always set as 0
    uint      npx;           // discritization in x direction
    uint      npy;           // discretization in y direction 
    uint      npz;           // discretization in y direction
 // select either unordered_map or map
    private:
    unordered_map<bitset<N>,value*, hash_override<N>>mesh;   
    map<bitset<N>,value*,  compare_override<N>>mesh;    
    unordered_map<bitset<N>,int> refinelist;   
    unordered_map<bitset<N>,int> derefinelist;   
  
    public:
    Tree(real *length,real *coords,uint nx,uint ny,uint nz); 
}
template <size_t N, typename value>
class Voxel : public Tree<N, value>
{
    private:
    uint maxlevel; //maximum level of refinement
    uint numMax;   //number of elements
    unordered_map<bitset<N>,value*>mesh;
  
    public:
    Voxel(real *length,real *coords) 
    // more member functions
    }  
\end{lstlisting}
\end{minipage}

The voxelization mesh has a similar structure as the solution mesh except that it need not be balanced. Ancestor of each element can be found using the bitwise operations. Therefore, the box that includes the corresponding geometric data can be identified quickly.
In flow computations around complex geometry, face normals of the solid geometry are also needed for the calculation of fluxes in finite-volume/discontinuous Galerkin methods or implementation of reconstruction schemes in the immersed boundary method. Voxelization is very useful in retrieving such geometric information. If one is only interested in mesh generation without any solution, the more efficient method of geometric encoding using bisection search is recommended, which we discuss next.

\subsection{Geometry encoding}
Each point of the geometry may be encoded in a similar fashion as the mesh elements. Given the coordinates of each point and the maximum level of adaptation, it is possible to generate a Morton code for each point at the highest level of adaptation. In other words, the finest level element that withholds the point is detected and saved in the class tree. This is basically equivalent to using the well-known bisection search in three-dimensional space and is only performed once regardless of the maximum level of adaptation. This way the search stage for tagging elements is eliminated altogether. 
Geometry encoding is illustrated for a one-dimension case in Algorithm \ref{alg:geometry_encode}. The three-dimensional case can be encoded by
using bit interleaving in each directions. In this algorithm, we accumulate true or false for each time we bisect a line and generate a code for a given point.

\begin{algorithm}
	\caption{Geometry encoding in one-dimension}
	\label{alg:geometry_encode}
	\begin{algorithmic}
		\State{define $maxLevel$ : maximum desired level of refinement which is a preset value}
		\State{define $Xmin$ : minimum x-coordinate of the computational domain}
		\State{define $Xmax$ : maximum x-coordinate of the computational domain}
		\State{define $xc$ : center coordinate}
		\State{define $xp$ : x-coordinate of a given points}
        \State{define $b$ : instance of std::bitset}
		\For{$ i= 0:maxLevel$}
		 \State{$xc=0.5\times(Xmax+Xmin)$}
		\If{$xp < xc$}
		\State{ $b(i)=false$}
				\State{ $Xmax=xc$}
		\Else
		 \State{ $b(i)=True$}		
		 		\State{ $Xmin=xc$}
		\EndIf
		\EndFor
	\end{algorithmic}
\end{algorithm}

%

\begin{figure}[h!]
	\centering
	\begin{tikzpicture}
	\begin{axis}[
    xlabel={Million Elements},
	ylabel={Total run time [s]},
	xmin=0, xmax=6,
	ymin=0, ymax=50,
	legend pos=north west,
	ymajorgrids=true,
	grid style=dashed,
	]

	\addplot[
	color=red,
	mark= triangle*,
	]
	coordinates {
	(.03,0.195)(0.12,0.25)(0.43,0.292)(1.2,0.84)(2.83,3.48)(5.87,12.4)
	};
	\addlegendentry{Geometric Encoding}
	
	\addplot[
	color=blue,
	mark=diamond*,
	]
	coordinates {

	(0.03,7.23)(0.12,7.6)(0.43,8.28)(1.2,8.47)(2.83,13.08)(5.87,45)
	};
	\addlegendentry{Voxel level 6}
	\end{axis}
	\end{tikzpicture}
	
	\caption{Effect of geometry encoding on overall execution time as a function of number of Cartesian mesh elements. Stanford bunny geometry with 88,754 surface triangles is used for this analysis }
	\label{fig:imp}
\end{figure}
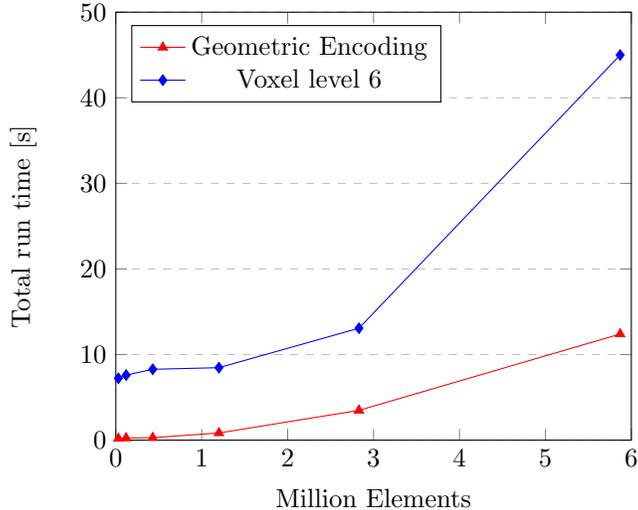

The expense to encode the geometry is $O(N)$, where $N$ is the number of triangular elements used to represent the solid geometry that we store as an STL file. The clear advantages of geometry encoding are that it does not grow as the solution mesh grows in size, and it is also cheaper in terms of time and memory than generating another coarse tree for voxelization. Figure \ref{fig:imp} shows the improvement in runtime due to geometry encoding and compare the runtime against the voxelization approach.

\section{Results}
\begin{figure}[h!]
  \begin{center}
 		\subfigure[Stanford bunny]{%
 			\label{fig:bunny}
 			\includegraphics[width=5.6cm,height=4.cm]{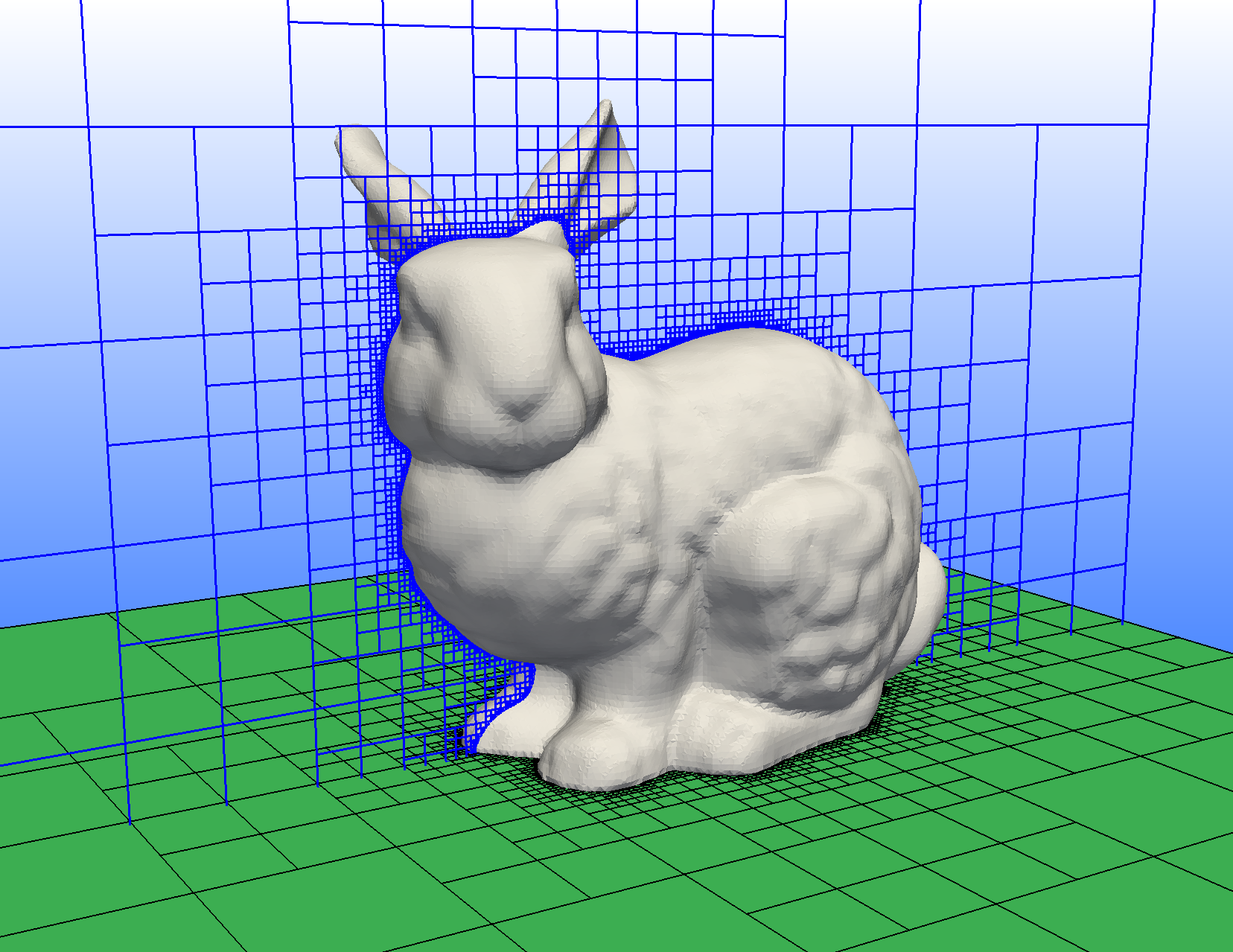}
 		}%
 		\subfigure[Sports car]{%
 			\label{fig:f16}
 			\includegraphics[width=5.6cm,height=4.cm]{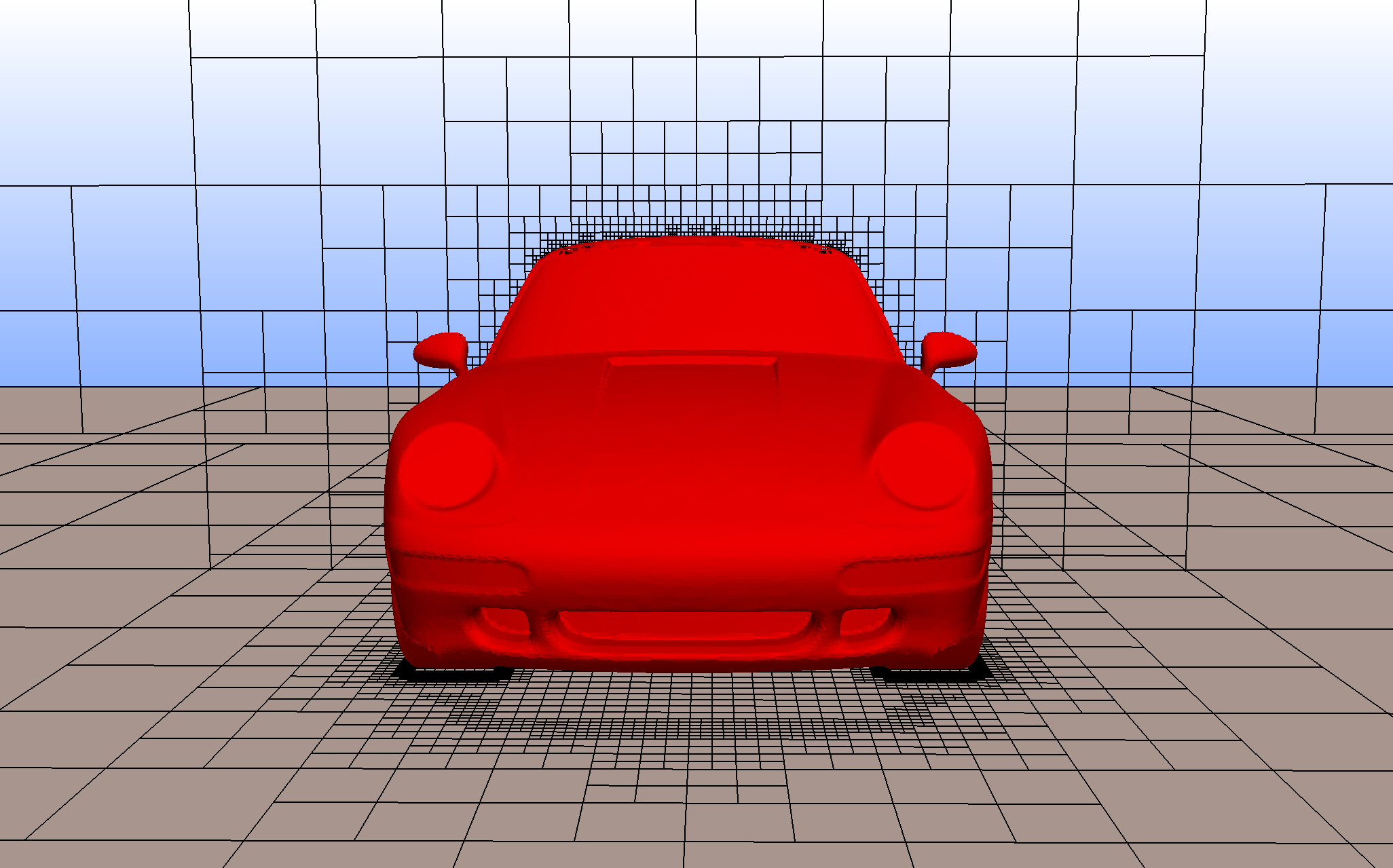}
 		}\\ 
 		\subfigure[Wind turbine]{%
 			\label{fig:turbine}
 			\includegraphics[width=5.6cm,height=4.cm]{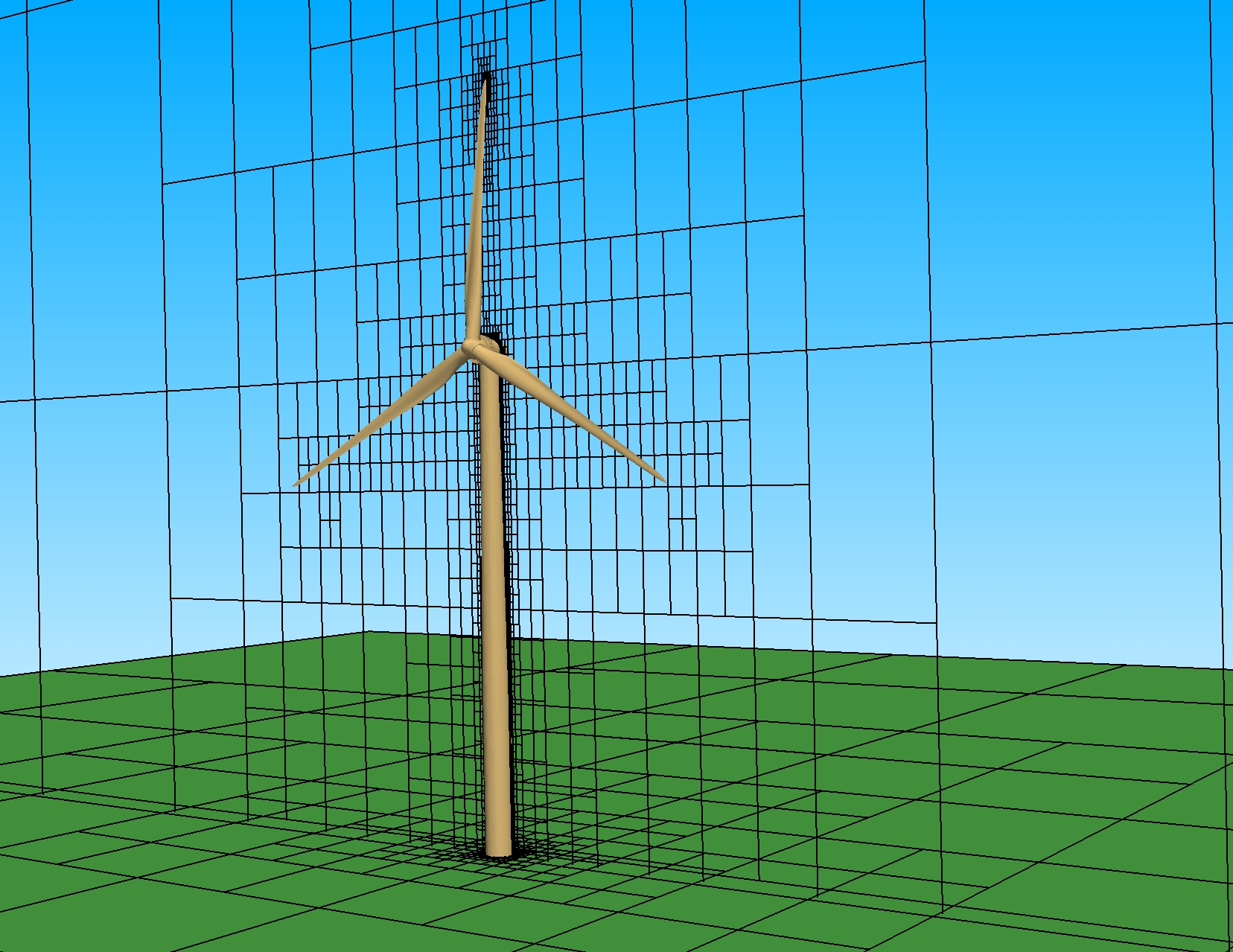}
 		} 
 		\subfigure[Mt. Hood, OR]{%
 			\label{fig:Mt}
 			\includegraphics[width=5.6cm,height=4.cm]{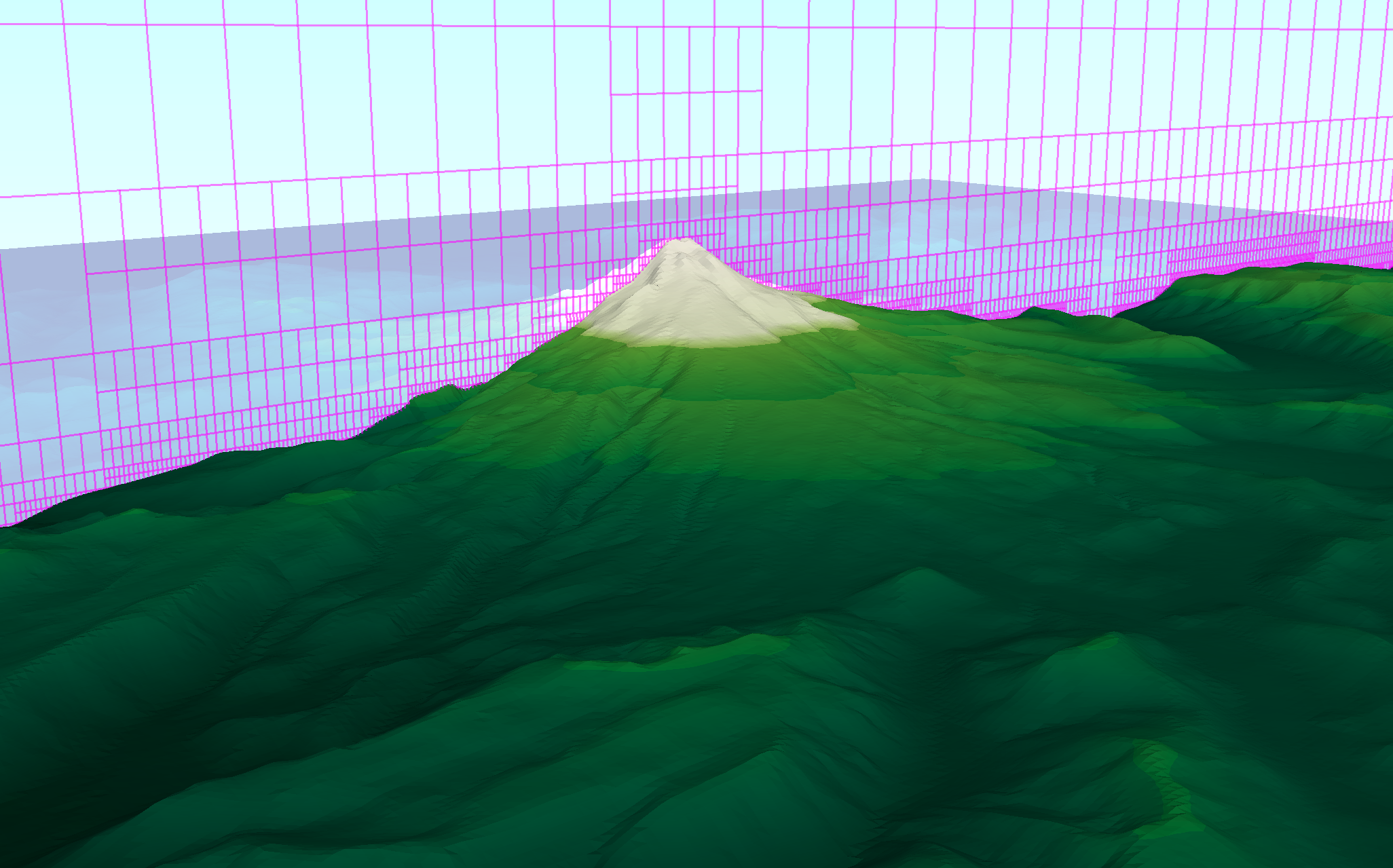}
 		}\\ 
 		
 \end{center}
 	\caption{%
 		Illustration of the 2:1 balanced Cartesian adaptive mesh generation for complex geometries from different fields.
 	}%
 	\label{fig:mesh}
\end{figure}

We present examples of 2:1 balanced octree generation for 
several STL geometries. The test cases include: Stanford bunny, which is a well-known test case from the computer graphics community; a sports car; a wind turbine; complex terrain of Mount Hood, OR. AMR for these geometries are presented in Fig. \ref{fig:mesh}, demonstrating the versatility of the current meshing capability.

\subsection{Performance analysis}
In this section, runtime and memory consumption of the hashed linear octree and binarized octree generation methods are investigated. We also demonstrate deep-level adaptation for the wind turbine geometry using the binarized octree method. Deep-level implies refinement levels that are greater than 21. All the calculations are carried out on a single compute-node of a Linux cluster with 64 GB of DDR3 1600 MHz memory and an Intel Xeon E5-2670 2.60 GHz processor.

The number of elements at different levels for the Stanford bunny and 
the sports car are presented in Table \ref{tab:Nelem}. These numbers are independent from the data structure utilized in the AMR algorithm. 

\begin{table}[]
	\centering
	\caption{Total mesh size for different adaptation levels}
	\begin{tabular}{ccccc}
		\\
		\toprule
		level & 10 & 12 & 14 & 16  \\
		\toprule
		 Bunny  (no. of triangles  =  88,754) \\   
		\midrule
		No. of elements [millions] & 0.43 & 2.83 & 10.52 & 23.96 \\
		\midrule
		Sports car  (no. of triangles  =  844,804)  \\
		\midrule
		No. of elements [millions]  & 0.11 & 1.84 & 16.18 & 74.09  \\
		\midrule
	\end{tabular}
	\label{tab:Nelem}
\end{table}

Here, using the two Stanford bunny and a sports car geometries, the run-time for binarized octree and hashed linear octree are presented for levels 10,
12, 14 and 16 in Tables \ref{tab:bunny_run_comp} and \ref{tab:porshce_run_comp}, respectively. The percentage of slow-down due to the use of a hash table is also presented for further clarification. For both of the geometries, the Z-compliant red-black tree is superior in terms of total runtime. 

\begin{table}[h!]
	\centering
	\caption{Comparison of total run-time. The Stanford bunny geometry is used for this analysis}
	\begin{tabular}{ccccc}
		\\
		\toprule
		level & 10 & 12 & 14 & 16  \\
		\toprule
        Bunny  (No. of triangles  =  88,754) \\   
		\midrule
		Z-curve compliant Binarized octree [seconds] & 3.3 & 21.14 & 69.75 & 145.87 \\
		\midrule
		Z-curve enhanced Hashed linear octree [seconds]  & 3.7 & 26.85 & 87.85 & 180.01  \\
		\midrule
	\end{tabular}
	\label{tab:bunny_run_comp}
 	
\end{table}

\begin{table}[h!]
	\centering
	\caption{Comparison of total run-time. The sports car geometry is used for this analysis}
	\begin{tabular}{ccccc}
		\\
		\toprule
		level & 10 & 12 & 14 & 16  \\
		\toprule
		Car  (No. of triangles  =  844,804) \\   
		\midrule
		Z-curve compliant binarized octree [seconds] & 1.58 & 19.68 & 164.56 & 696.54  \\
		\midrule
		Z-curve enhanced hashed linear octree [seconds]  & 1.64 & 24.73 & 260.88 & 753.58  \\
		\midrule
	\end{tabular}
	\label{tab:porshce_run_comp}
\end{table}

Memory consumption for the binarized octree and the hashed linear octree are presented for levels 12, 10, 14 and 16 in Tables \ref{tab:bunny_memory} and \ref{tab:porsche_memory} for the same geometries discussed in the previous section. Valgrind is used for memory profiling \cite{nethercote2007shadow}.According to the results, hashed linear octree generation method use less memory than the binarized octree method, but the difference in memory usage is not substantial enough to cause any concern. 

\begin{table}[h!]
	\centering
	\caption{Comparison of memory consumption. The Stanford bunny geometry is used for this analysis.}
	\begin{tabular}{ccccc}
		\\
		\toprule
		level & 10 & 12 & 14 & 16  \\
    	\toprule
		Bunny (Ntriangles  =  88,754) \\ 
		\midrule
		Z-curve compliant binarized octree [MB] & 61.4  & 281.8 & 792.9 & 1500  \\
		\midrule
		Z-curve enhanced  hashed linear octree [MB] & 59.6  & 261.7  & 675.5 & 1300  \\
		\midrule
		extra memory usage [\%]  &3.0 & 7.6 & 17.4 & 15.4 \\
		\midrule
	\end{tabular}
	\label{tab:bunny_memory}
\end{table}

\begin{table}[h!]
	\centering
	\caption{Comparison of memory consumption. The sports car geometry is used for this analysis.}
	\begin{tabular}{ccccc}
		\\
		\toprule
		level & 10 & 12 & 14 & 16  \\
		\toprule
		Car (No. of triangles  =  844,804) \\ 
		\midrule
		Z-curve compliant binarized octree [MB] & 106.3  & 341.1 & 1800 & 5800  \\
		\midrule
		Z-curve enhanced hashed linear octree [MB] & 106.3  & 305.7   & 2000 & 5100  \\
		\midrule
		extra memory usage [\%]  &0.0 & 11.6 & -11.1 & 13.7\\
		\midrule
	\end{tabular}
	\label{tab:porsche_memory}
\end{table}

Unstructured mesh approaches use much more memory than their tree-based AMR counterparts. The total run-time using puAMR \cite{hasbestan2017parallel} on a single processor is compared against the performance of the binarized octree generation method in Table \ref{tab:tree_graph}. In this comparison, we are not able to go beyond 16 levels of adaption, because the unstructured approach runs out of computer memory. The other important factor to mention here is that in the unstructured approach, the search time for tagging elements for refinement is much longer than the binarized octree generation that use geometric encoding for the search algorithm. The slow performance of the unstructured approach is expected and we include it in this study to highlight the superior performance of tree-based approaches for AMR.

\begin{table}[h!]
	\centering
	\caption{Comparison of runtime for binarized octree against the graph-based unstructured mesh approach. Stanford bunny geometry is used for this analysis.}
	\label{tab:tree_graph}
	\begin{tabular}{cccccc}
		\toprule
		level & 7 & 8 & 9 & 10 & 11\\
		\midrule
		mesh size & 8163 & 31242 & 118357 & 437,557 & 1,199,290  \\
		\midrule
		Z-curve compliant \\ Binarized octree[seconds] & 0.16 & 0.21 & 0.438  & 2.57 & 4.71  \\ 
		\midrule
		Graph-based[seconds] & 3.16 &  4.71 & 9.02 & 32.69 & 615.12 \\
		\midrule 
	\end{tabular}
\end{table}

Table \ref{tab:deep} demonstrates the deep-level adaption that is inherently feasible with the binarized octree generation approach. We use the wind turbine geometry with 70,406 surface triangles. The mesh around this geometry is presented in Fig. \ref{fig:turbine}. 


\begin{table}[h!]
	\centering
	\caption{Number of mesh elements as a function of adaptation level for a binarized octree generation method. Wind turbine geometry with 70,406 surface triangles is used. Note that hashed linear octree method limits the maximum level of adaption to 21 on 64-bit systems.}
	\label{tab:deep}
	\begin{tabular}{ccccc}
		\\
		\toprule
		level & 20 & 21 & 22 & 23 \\
		\midrule
		mesh size & 1,246,330 & 2,925,973 & 6,007,583 & 10,714,012\\
	\end{tabular}
\end{table}

\section{Conclusions}
\label{sec:conclusions}
There are various algorithms for adaptive mesh refinement. Performance of these different approaches can vary greatly. In the present work, we proposed the binarized octree generation method by combining Morton encoding with a red-black tree, which is special kind of self balancing binary tree. In our computer implementation, we benefited from the C++ STL \texttt{bitset} \texttt{Set/Map} classes. The C++ STL \texttt{bitset} enables us to use an arbitrary length for the bit representation regardless of the hardware architecture. Additionally, the red-black tree operates on comparison of bits only. Because of these two features, we are able generate meshes with arbitrary levels of adaptation on a single processor. On the contrary, the hashed linear octree generation method is restricted to 21 levels for 64-bit systems, beyond which an overflow is inevitable because of conversion to an integer number in the hash function. However, one can relax the Z-order curve compliance and generate arbitrary orders of adaptation at the expense of data-locality and loss of runtime performance. 

Another advantage of using the red-black tree in octree generation is that it keeps its elements sorted by default and preserves the Z-order curve exactly. In the hashed linear octree generation method, the order can get broken because of collisions in the hash table. An important implication of this strict compliance with the Z-order curve is that the generated mesh is ready to be partitioned for parallel computations without any need for sorting its elements.

We exploited the bitwise representation of an octree and developed techniques to extract neighborhood connectivity from the Morton codes without explicitly storing the connectivity information. Additionally, we developed a geometry encoding technique to rapidly tag the elements for refinement.   

Our results show that the binarized octree technique is faster than the hashed linear octrree technique at the expense of a slight increase in memory usage. The current work is released as open-source code which includes both the binarized and the hashed linear octree generation methods \cite{rbAMR}.

\section*{Acknowledgments}
This material is based upon work supported by the National Science Foundation under Grant No. 1440638 and Army Research Office under Grant No. W911NF-17-1-0564.

\section*{References}


\end{document}